\documentclass[onecolumn,unpublished,a4paper,accepted=2025-01-06]{quantumarticle}
\pdfoutput=1
\usepackage[utf8]{inputenc}
\usepackage[backend=biber,style=alphabetic,maxbibnames=999]{biblatex}
\usepackage[colorlinks]{hyperref}
\usepackage{graphicx}
\usepackage{amsfonts}
\usepackage{amsmath}
\usepackage{mathtools}
\usepackage{cleveref}
\usepackage{enumitem}
\usepackage{algorithm}
\usepackage{algpseudocode}
\usepackage{caption}
\usepackage{subcaption}
\usepackage{siunitx}

\newcommand{\fig}[1]{\hyperref[fig:#1]{Figure~\ref*{fig:#1}}}
\newcommand{\eq}[1]{\hyperref[eq:#1]{Equation~\ref*{eq:#1}}}
\newcommand{\tab}[1]{\hyperref[tab:#1]{Table~\ref*{tab:#1}}}
\newcommand{\sect}[1]{\hyperref[sec:#1]{Section~\ref*{sec:#1}}}
\newcommand{\app}[1]{\hyperref[app:#1]{Appendix~\ref*{app:#1}}}
\newcommand{\theoremref}[1]{\hyperref[theorem:#1]{Theorem~\ref*{theorem:#1}}}
\newcommand{\definitionref}[1]{\hyperref[definition:#1]{Definition~\ref*{definition:#1}}}

\bibliography{references}

\begin{document}
\renewcommand\thesection{\arabic{section}}  %

\title{Sparse Blossom: correcting a million errors per core second with minimum-weight matching}

\author{Oscar Higgott}
\affiliation{Google Quantum AI, Santa Barbara, California 93117, USA}
\affiliation{Department of Physics \& Astronomy, University College London, WC1E 6BT London, United Kingdom}
\author{Craig Gidney}
\affiliation{Google Quantum AI, Santa Barbara, California 93117, USA}
\date{January 14, 2025}

\maketitle

\begin{abstract}
In this work, we introduce a fast implementation of the minimum-weight perfect matching (MWPM) decoder, the most widely used decoder for several important families of quantum error correcting codes, including surface codes. Our algorithm, which we call sparse blossom, is a variant of the blossom algorithm which directly solves the decoding problem relevant to quantum error correction. Sparse blossom avoids the need for all-to-all Dijkstra searches, common amongst MWPM decoder implementations. For 0.1\% circuit-level depolarising noise, sparse blossom processes syndrome data in both X and Z bases of distance-17 surface code circuits in less than one microsecond per round of syndrome extraction on a single core, which matches the rate at which syndrome data is generated by superconducting quantum computers. Our implementation is open-source, and has been released in version 2 of the PyMatching library.
\end{abstract}

\emph{
The source code for our implementation of sparse blossom in PyMatching version 2 can be found on github at \url{https://github.com/oscarhiggott/PyMatching}.
PyMatching is also available as a Python 3 pypi package installed via ``}\texttt{pip install pymatching}\emph{".}

\section{Introduction}

A surface code superconducting quantum computer with a million physical qubits will generate measurement data at a rate of around 1 terabit per second.
This data must be processed at least as fast as it is generated by a \textit{decoder}, the classical software used to predict which errors may have occurred, to prevent a backlog of data that grows exponentially in the $T$-gate depth of the computation~\cite{terhal2015quantum}.
Moreover, fast decoders are important not only for operating a quantum computer, but also as a software tool for researching quantum error correction protocols.
Estimating the resource requirements of a quantum error correction protocol (e.g.~estimating the teraquop footprint below threshold~\cite{Gidney2021faulttolerant}) can require the use of direct Monte Carlo sampling to estimate the probability of extremely rare events.
These analyses can be prohibitively expensive without access to a fast decoder, capable of processing millions of shots from circuits containing $\approx 10^5$ gates in a reasonable time-frame.

Many decoders have been developed for the surface code, the oldest and most popular of which is the minimum-weight perfect matching (MWPM) decoder~\cite{dennis2002topological}, which is also the focus of this work.
The MWPM decoder maps the decoding problem onto a graphical problem by decomposing the error model into $X$-type and $Z$-type Pauli errors~\cite{dennis2002topological}.
This graphical problem can then be solved with the help of Edmonds' blossom algorithm for finding a minimum-weight perfect matching in a graph~\cite{edmonds1965paths,edmonds1965maximum}.
A naive implementation of the MWPM decoder has a worst-case complexity in the number of nodes $N$ in the graph of $O(N^3\log(N))$, with the expected running time for typical instances found empirically to be roughly $O(N^2)$~\cite{higgott2022pymatching}.
Approximations and optimisations of the MWPM decoder have led to significantly improved expected running times~\cite{fowler2012towards,fowler2013minimum, higgott2022pymatching}. In particular, \citeauthor{fowler2013minimum} proposed a MWPM decoder with average $O(1)$ parallel running time~\cite{fowler2013minimum}.
However, published implementations of MWPM decoders have not demonstrated speeds fast enough for real-time decoding at scale.

There have been several alternatives to the MWPM decoder proposed in the literature.
The Union-Find decoder has an almost-linear worst-case running time~\cite{Delfosse2021almostlineartime, huang2020fault}, and fast hardware implementations have been proposed~\cite{das2020scalable} and implemented~\cite{liyanage2023scalable}.
The Union-Find decoder is slightly less accurate than, and can be seen as an approximation of, the MWPM decoder~\cite{wu2022interpretation}.
Maximum-likelihood decoders can achieve a higher accuracy than the MWPM decoder~\cite{bravyi2014efficient,higgott2022fragile, sundaresan2022matching} but have high computational complexity, rendering them impractical for real-time decoding.
Other decoders, such as correlated MWPM~\cite{fowler2013optimal}, belief-matching~\cite{higgott2022fragile} and neural network~\cite{torlai2017neural,meinerz2022scalable} decoders can achieve higher accuracy than MWPM with a much more modest increase in running time.
While there has been progress in the development of open-source software packages for decoding surface codes~\cite{higgott2022pymatching,qecsim}, these tools are much slower than stabilizer circuit simulators~\cite{Gidney2021stimfaststabilizer}, and have therefore been a bottleneck in surface code simulations. This is perhaps one of the reasons why numerical studies of error correcting codes have often focused on estimating \textit{thresholds} (which require decoding fewer shots), instead of resource overheads (which are more practically useful for making comparisons).

In this work, we introduce a new implementation of the MWPM decoder.
The algorithm we introduce, sparse blossom, is a variant of the blossom algorithm which is conceptually similar to the approach taken in Refs.~\cite{fowler2012towards,fowler2012timing_analysis, fowler2013minimum}, in that it solves the MWPM decoding problem directly on the detector graph, rather than naively breaking up the problem into multiple sequential steps and solving the traditional MWPM graph theory problem as a separate subroutine.
This avoids the all-to-all Dijkstra searches often used in implementations of the MWPM decoder.
Our implementation is orders of magnitude faster than alternative available tools, and can decode both $X$ and $Z$ bases of a distance-17 surface code circuit (for $0.1\%$ circuit-noise) in under one microsecond per round on a single core, matching the rate at which syndrome data is generated on a superconducting quantum processor.
At distance 29 with the same noise model (more than sufficient to achieve $10^{-12}$ logical error rates), PyMatching takes 3.5 microseconds per round to decode on a single core.
These results suggest that our sparse blossom algorithm is fast enough for real-time decoding of superconducting quantum computers at scale; a real-time implementation is likely achievable through parallelisation across multiple cores, and by adding support for decoding a stream, rather than a batch, of syndrome data.
Our implementation of sparse blossom has been released in version 2 of the PyMatching Python package, and can be combined with Stim~\cite{Gidney2021stimfaststabilizer} to run simulations in minutes on a laptop that previously would have taken hours on a high-performance computing cluster.

In impressive parallel independent work, Yue Wu has also developed a new implementation of the blossom algorithm called fusion blossom~\cite{wu2023fusion}, available at \cite{wu2022fusionblossom}.
The conceptual similarity with our approach is that fusion blossom also solves the MWPM decoding problem directly on the detector graph.
However, there are many differences in the details of our respective implementations; for example, fusion blossom explores the graph in a similar way to how clusters are grown in union-find, whereas our approach grows exploratory regions uniformly, managed by a global priority queue.
While our approach has faster single-core performance, fusion blossom also supports parallel execution of the algorithm itself, which can be used to achieve faster processing speeds for individual decoding instances.
When used for error correction simulations, we note that sparse blossom is already trivially parallelisable by splitting the simulation into batches of shots, and processing each batch on a separate core.
However, parallelisation of the decoder itself is important for real-time decoding, to prevent an exponentially increasing backlog of data building up within a single computation~\cite{terhal2015quantum}, or to avoid the polynomial slowdown imposed by relying on parallel window decoding instead~\cite{skoric2022parallel,tan2022scalable}.
Therefore, future work could explore combining sparse blossom with the techniques for parallelisation introduced in fusion blossom.

The paper is structured as follows. In \sect{preliminaries} we give background on the decoding problem we are interested in and give an overview of the blossom algorithm. In \sect{sparse_blossom} we explain our algorithm, sparse blossom, before describing the data structures we use for our implementation in \sect{data_structures}. In \sect{expected_running_time} we analyse the running time of sparse blossom, and in \sect{computational_results} we benchmark its decoding time, before concluding in \sect{conclusion}.

\section{Preliminaries}\label{sec:preliminaries}

\subsection{Detectors and observables}

A \textit{detector} is a parity of measurement outcome bits in a quantum error correction circuit that is deterministic in the absence of errors.
The outcome of a detector measurement is 1 if the observed parity differs from the expected parity for a noiseless circuit, and is 0 otherwise.
We say that a Pauli error $P$ \textit{flips} detector $D$ if including $P$ in the circuit changes the outcome of $D$, and a \textit{detection event} is a detector with outcome 1.
We define a \textit{logical observable} to be a linear combination of measurement bits, whose outcome instead corresponds to the measurement of a logical Pauli operator.

We define an independent error model to be a set of $m$ independent error mechanisms, where error mechanism $i$ occurs with probability $\mathbf{p}[i]$ (where $\mathbf{p}\in\mathbb{R}^m$ is a vector of priors), and flips a set of detectors and observables.
An independent error model can be used to represent circuit-level depolarising noise exactly, and is a good approximation of many commonly considered error models~\cite{Chao2020optimizationof, gidney2020decorrelated}.
It can be useful to describe an error model in an error correction circuit using two binary parity check matrices: a \textit{detector check matrix} $H\in \mathbb{F}_2^{n\times m}$ and a \textit{logical observable matrix} $L\in \mathbb{F}_2^{n_l\times m}$.
We set $H[i, j]=1$ if detector $i$ is flipped by error mechanism $j$, and $H[i, j]=0$ otherwise.
Likewise, we set $L[i, j]=1$ if logical observable $i$ 
is flipped by error mechanism $j$, and $L[i, j]=0$ otherwise.
By describing the error model this way, each error mechanism is defined by which detectors and observables it flips, rather than by its Pauli type and location in the circuit.
From a stabilizer circuit and Pauli noise model, we can construct $H$ and $L$ efficiently by propagating Pauli errors through the circuit to see which detectors and observables they flip.
Each prior $\mathbf{p}[i]$ is then computed by summing over the probabilities of all the Pauli errors that flip the same set of detectors or observables (or more precisely, these equivalent error mechanisms are independent, and we compute the probability that an odd number occurred).
This is essentially what the error analysis tools do in Stim, where a detector error model (automatically constructed from a Stim circuit) captures the information contained in $H$, $L$ and $\mathbf{p}$~\cite{Gidney2021stimfaststabilizer}.

We can represent an error (a set of error mechanisms) by a vector $\mathbf{e}\in\mathbb{F}_2^{m}$.
The \textit{syndrome} of $\mathbf{e}$ is the outcome of detector measurements, given by $\mathbf{s}=H\mathbf{e}$.
The detection events are the detectors corresponding to the non-zero elements of $\mathbf{s}$.
An undetectable logical error is an error in $B\coloneqq \{\mathbf{e}\in\mathbb{F}_2^{m}\mid \mathbf{e}\in\ker{H}, \mathbf{e}\not\in\ker{L}\}$, and the distance of the circuit is $d=\min_{\mathbf{e}\in B}|\mathbf{e}|$, where $|\mathbf{e}|$ is the Hamming weight of $\mathbf{e}$.
Given a syndrome $\mathbf{s}=H\mathbf{e}$ of some error $\mathbf{e}\in\mathbb{F}_2^{m}$, as well as knowledge of $H$, $L$ and $\mathbf{p}$, a decoder makes a prediction $\mathbf{c}\in\mathbb{F}_2^{m}$ of an error satisfying $H\mathbf{c}=H\mathbf{e}$ and succeeds if $L(\mathbf{c}\oplus\mathbf{e})=0$. Note that sometimes a decoder need only output the logical prediction $L\mathbf{c}$ (succeeding if it matches $L\mathbf{e}$), as we discuss later.

For a given circuit, the choice of detectors is not unique.
For example, if we redefine any detector $D_i$ as $D_i\coloneqq D_i\oplus D_j$ then we still have a valid choice of detectors, and this change corresponds to adding row $j$ to row $i$ of $H$.
Since two errors $\mathbf{e}_k,\mathbf{e}_l\in\mathbb{F}_2^{m}$ are distinguishable if and only if $\mathbf{e}_k\oplus \mathbf{e}_l\notin \ker{H}$, two detector check matrices can distinguish the same errors if they have the same kernel~\cite{derks2024designingfaulttolerantcircuitsusing}.
While row operations on $H$ do not affect which errors can be distinguished, it is often necessary that the choice of basis for the detectors has a particular structure in order for a given decoding algorithm to be applicable (as discussed shortly in \sect{detector_graphs}).

Similarly, the choice of logical observables is also not unique.
Since detector outcomes are deterministically zero in the absence of noise, we can add any linear combination of detectors to a logical observable without effecting its expectation value for a noiseless circuit.
In terms of check matrices, this change corresponds to defining some new logical observables matrix $L^\prime$ where we have added some linear combinations of the rows of $H$ to each row of the original logical observables matrix $L$:
\begin{equation}
    L^\prime \coloneqq L \oplus AH.
\end{equation}
Here $A\in \mathbb{F}_2^{n_l\times n}$ is an arbitrary binary matrix of dimensions $n_l\times n$.
If we use a decoder that is guaranteed to provide a prediction consistent with the syndrome, i.e.~$H(\mathbf{c}\oplus \mathbf{e})=0$, then adding detectors to each observable will not change whether or not the decoder succeeds, since $L^\prime (\mathbf{c}\oplus \mathbf{e}) = L (\mathbf{c}\oplus \mathbf{e})$.

In \app{repetition_code_example}, we show how the detectors, observables and the matrices $H$ and $L$ are defined for a small example of a distance 2 repetition code circuit.

\subsection{Detector graphs}\label{sec:detector_graphs}

In this work, we will restrict our attention to \textit{graphlike error models}, defined to be independent error models for which each error mechanism flips at most two detectors ($H$ has column weight at most two).
Graphlike error models can be used to approximate common noise models for many important classes of quantum error correction codes including surface codes~\cite{dennis2002topological}, for which $X$-type and $Z$-type Pauli errors are both graphlike.
Many related code families also have graphlike error models, such as repetition codes, 2D hyperbolic codes~\cite{breuckmann2016constructions}, some 2D subsystem codes~\cite{bravyi2013subsystemsurfacecodesthreequbit, Suchara2011constructions, bacon2006operator, higgott2021subsystem} and Floquet codes~\cite{Hastings2021dynamically, Gidney2021faulttolerant}, amongst others.
Color codes~\cite{bombin2006topological} can be decoded by using a mapping to a graphlike error model as a subroutine~\cite{Kubica2023efficientcolorcode, gidney2023newcircuitsopensource}, and generalizing this approach to decode more general codes is an active area of research~\cite{brown2022conservation}.

We can represent a graphlike error model using a detector graph $\mathcal{G}=(\mathcal{V}, \mathcal{E})$, also called a matching graph or decoding graph in the literature.
Each node $v\in\mathcal{V}$ corresponds to a detector (a detector node, a row of $H$).
Each edge $e\in\mathcal{E}$ is a set of detector nodes of cardinality one or two representing an error mechanism that flips this set of detectors (a column of $H$).
We can decompose the edge set as $\mathcal{E}=\mathcal{E}_1\cup\mathcal{E}_2$ where $\forall e\in\mathcal{E}_1: |e|=1$ and $\forall e\in\mathcal{E}_2: |e|=2$.
A regular edge $(u, v)\in \mathcal{E}_2$ flips a pair of detectors $u, v\in\mathcal{V}$, whereas a \textit{half-edge} $(u,)\in\mathcal{E}_1$ flips a single detector $u\in\mathcal{V}$.
For a half-edge $(u,)\in\mathcal{E}_1$ we sometimes say that $u$ is connected to the boundary and use the notation $(u,v_b)$, where $v_b$ is a virtual boundary node (which does not correspond to any detector).
Therefore, when we refer to an edge $(u,v)\in\mathcal{E}$ it is assumed that $u$ is a node and $v$ is either a node or the boundary node $v_b$.
Each edge $e_i\in\mathcal{E}$ is assigned a weight $w(e_i)=\log((1-\mathbf{p}[i])/\mathbf{p}[i])$, and recall that $\mathbf{p}[i]$ is the probability that error mechanism $i$ occurs.
We also define an edge weights vector $\mathbf{w}\in \mathbb{R}^{|\mathcal{E}|}$ for which $\mathbf{w}[i]=w(e_i)$.
We also label each edge $e_i=(u,v)\in\mathcal{E}$ with the set of logical observables that are flipped by the error mechanism, which we denote either by $l(e_i)$ or $l(u,v)$.
We use $x\oplus y$ to denote the symmetric difference of sets $x$ and $y$.
For example, $l(e_1)\oplus l(e_2)$ is the set of logical observables flipped when the error mechanisms 1 and 2 are both flipped.
We define the distance $D(u, v)$ between two nodes $u$ and $v$ in the detector graph to be the length of the shortest path between them.
We give an example of a detector graph $\mathcal{G}$ for a repetition code circuit in \app{repetition_code_example}.

\subsection{The minimum-weight perfect matching decoder}

From now on, we will assume that we have an independent graphlike error model with check matrix $H$, logical observables matrix $L$, priors vector $\mathbf{p}$, as well as the corresponding detector graph $\mathcal{G}$ with edge weights vector $\mathbf{w}$.
Given some error $\mathbf{e}\in\mathbb{F}_2^{m}$ sampled from the graphlike error model, with the observed syndrome $\mathbf{s}=H\mathbf{e}$, the minimum-weight perfect matching (MWPM) decoder finds the \textit{most probable physical error} consistent with the syndrome.
In other words, for a graphlike error model it finds a physical error $\mathbf{c}\in\mathbb{F}_2^{m}$ satisfying $H\mathbf{c}=\mathbf{s}$ that has maximum prior probability $P(\mathbf{c})$.
For our error model the prior probability $P(\mathbf{e})$ of an error $\mathbf{e}$ is
\begin{equation}
    P(\mathbf{e})=\prod_i(1-\mathbf{p}[i])^{(1-\mathbf{e}[i])}\mathbf{p}[i]^{\mathbf{e}[i]}=\prod_i(1-\mathbf{p}[i])\prod_i\left(\frac{\mathbf{p}[i]}{1-\mathbf{p}[i]}\right)^{\mathbf{e}[i]}.
\end{equation}
Equivalently we seek to maximise $\log(P(\mathbf{c}))=C-\sum_i \mathbf{w}[i]\mathbf{c}[i]$, where here $C\coloneqq \sum_i \log(1-\mathbf{p}[i])$ is a constant and we recall that the edge weight is defined as $\mathbf{w}[i]\coloneqq \log((1-\mathbf{p}[i])/\mathbf{p}[i])$. 
This corresponds to finding an error $\mathbf{c}\in \mathbb{F}_2^m$ satisfying $H\mathbf{c}=\mathbf{s}$ that minimises $\sum_i \mathbf{w}[i]\mathbf{c}[i]$, the sum of the weights of the corresponding edges in $\mathcal{G}$.
We note that a maximum likelihood decoder instead finds the most probable \textit{logical} error for any error model (graphlike or not), however maximum likelihood decoding is in general computationally inefficient.

Despite the MWPM decoder's name, the problem it solves, defined above, does not correspond to finding a MWPM in $\mathcal{G}$, but instead corresponds to solving a variation of the MWPM problem, which we refer to as the minimum weight \textit{embedded} matching (MWEM) problem.
Let us first define the traditional MWPM problem for a graph $G=(V, E)$.
Here $G$ is a weighted graph, where each edge $(u, v)\in E$ is a pair of nodes $u, v\in V$ and, unlike detector graphs, there are no half-edges.
Each edge is assigned a weight $w(e)\in\mathbb{R}$.
A perfect matching $M\subseteq E$ is a subset of edges such that each node $u\in V$ is incident to exactly one edge $(u, v)\in M$.
For each $(u, v)\in M$ we say that $u$ is matched to $v$, and vice versa.
A MWPM is a perfect matching that has minimum weight $\sum_{e\in M}w(e)$.
Clearly, not every graph has a perfect matching (a simple necessary condition is that $|V|$ must be even; a necessary \textit{and sufficient} condition is provided by Tutte's theorem~\cite{tutte1947factorization}), and a graph may have more than one perfect matching of minimum weight.

\begin{figure}
    \centering
    \includegraphics[width=\columnwidth]{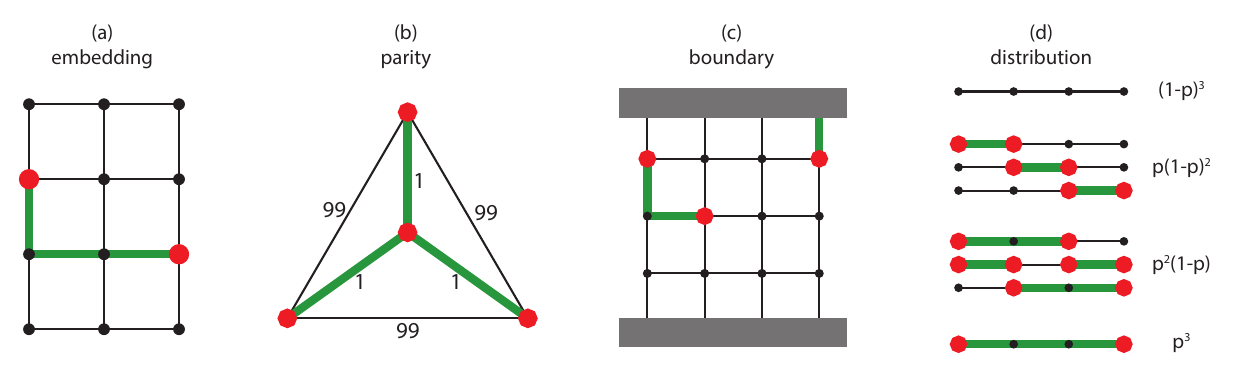}
    \caption{
    Key differences between the quantum decoding problem solved by PyMatching and the minimum weight perfect matching problem.
    In the usual MWPM problem, all nodes must be matched and they are matched using a disjoint set of edges.
    In the decoding problem, (a) only a subset of nodes is excited, only these nodes need to be matched, and (b) the edge set used to match them is not required to be disjoint.
    The excited nodes are matched by finding an edge set where excited nodes have an odd number of neighbors in the edge set, non-excited nodes have an even number of neighbors in the edge set, and (c) there may be boundary nodes that can have any number of neighbors in the edge set.
    (d) The expected distribution of excited nodes is not uniform.
    It is generated by sampling edges, where each edge is independently included with some probability, and then exciting any nodes that have an odd number of neighbors in the sampled edge set.
    This results in it being exponentially unlikely to see large distances between excited nodes at low error rates, which has major implications on the expected runtime of the algorithm (see \sect{expected_running_time}).
    }
    \label{fig:difference_from_mwpm}
\end{figure}

Given a detector graph $\mathcal{G}=(\mathcal{V}, \mathcal{E})$ with vertex set $\mathcal{V}$, and a set of detection events (highlighted nodes) $\mathcal{D}\subseteq \mathcal{V}$, we define an \textit{embedded} matching of $\mathcal{D}$ in $\mathcal{G}$ to be a set of edges $M_E\subseteq \mathcal{E}$ such that every node in $\mathcal{D}$ is incident to an \textit{odd} number of edges in $M_E$, and every node in $\mathcal{V}\setminus \mathcal{D}$ is incident to an \textit{even} number of edges in $M_E$.
The minimum-weight embedded matching is an embedded matching that has minimum weight $\sum_{e\in M_E}w(e)$.
We note that the minimum-weight embedded matching problem for a standard graph (containing no half-edges) is known as the minimum-weight $T$-join problem in the field of combinatorial optimisation (for $T\equiv \mathcal{D}$)~\cite{edmonds1973matching,korte2011combinatorial}.
The key differences between the minimum-weight embedded matching problem we are interested in for error correction, and the traditional MWPM problem, are shown in \fig{difference_from_mwpm}.

The decoder does not need to output the set of edges directly, but rather outputs $L\mathbf{c}$, a prediction of which logical observable measurements were flipped.
Our decoder has succeeded if it correctly predicts which observables were flipped, i.e.~if $L\mathbf{c}=L\mathbf{e}$.
In other words, we apply a correction at the logical level rather than at the physical level, which is equivalent since $L(\mathbf{c} \oplus \mathbf{e})=L\mathbf{c} \oplus L\mathbf{e}$.
This output is generally much more sparse: for example, in a surface code memory experiment the prediction is simply a single bit, predicting whether or not the logical $X$ (or $Z$) observable measurement outcome was flipped by the error.
As we will see later, predicting logical observables $L\mathbf{c}$ rather than the full physical error $\mathbf{c}$ leads to some useful optimizations.
Note that PyMatching does also support returning the full physical error (e.g.~a unique observable can be assigned to each edge), but we apply these additional optimizations when the number of logical observables is small (up to 64 observables).

Note that the edge weight $\mathbf{w}[i]=\log((1-\mathbf{p}[i])/\mathbf{p}[i])$ is negative when $\mathbf{p}[i]>0.5$.
Our sparse blossom algorithm instead assumes that edge weights are non-negative.
Fortunately, it is straightforward to decode a syndrome $\mathbf{s}$ for a detector graph $\mathcal{G}$ containing negative edge weights by using efficient pre- and post-processing to instead decode a modified syndrome $\mathbf{s}^\prime$ using a set of adjusted edge weights $\mathbf{v}$ containing only non-negative edge weights.
This procedure is used to handle negative edge weights in PyMatching and is explained in \app{negative_edge_weights}.
However, from now on we will assume, without loss of generality, that we have a graph containing only non-negative edge weights.

\subsection{Connection between minimum-weight perfect and embedded matching}\label{sec:mwem_using_mwpm}

Although minimum-weight embedded and perfect matchings are different problems, there is a close connection between them.
In particular, by using a polynomial-time algorithm for solving the MWPM problem (e.g.~the blossom algorithm), we can obtain a polynomial-time algorithm for solving the MWEM problem via a reduction.
We will now describe this reduction for the case that the detector graph $\mathcal{G}$ has no boundary, in which case the MWEM problem is equivalent to the minimum-weight $T$-join problem (see \cite{edmonds1973matching,korte2011combinatorial}).
This reduction was used by Edmonds and Johnson for their polynomial-time algorithm for solving the Chinese postman problem~\cite{edmonds1973matching}.
The boundary can also be handled with a small modification (e.g.~see \cite{fowler2013minimum}).

Given a detector graph $\mathcal{G}=(\mathcal{V}, \mathcal{E})$ with non-negative edge weights (and which, for now, we assume has no boundary, i.e.~$\mathcal{E}=\mathcal{E}_2$), and given a set of detection events $\mathcal{D}\subseteq \mathcal{V}$, we define the \textit{path graph} $\bar{\mathcal{G}}[\mathcal{D}]=(\mathcal{D}, \bar{\mathcal{E}})$ to be the complete graph on the vertices $\mathcal{D}$ for which each edge $(u,v)\in\bar{\mathcal{E}}$ is assigned a weight equal to the distance $D(u,v)$ between $u$ and $v$ in $\mathcal{G}$.
Here the distance $D(u,v)$ is the length of the shortest path between $u$ and $v$ in $\mathcal{G}$, where here the length of the path is the sum of the weights of the edges it contains.
In other words, the path graph $\bar{\mathcal{G}}[\mathcal{D}]$ is the subgraph of the metric closure of $\mathcal{G}$ induced by the vertices $\mathcal{D}$.
A MWEM $M$ of $\mathcal{D}$ in $\mathcal{G}$ can be found efficiently using the following three steps:
\begin{enumerate}
    \item Construct the path graph $\bar{\mathcal{G}}[\mathcal{D}]$ using Dijkstra's algorithm.
    \item Find the minimum-weight perfect matching $\bar{M}\subseteq \bar{\mathcal{E}}$ in $\bar{\mathcal{G}}[\mathcal{D}]$ using the blossom algorithm.
    \item Use $\bar{M}$ and Dijkstra's algorithm to construct the MWEM: $M\coloneqq \bigoplus_{(u,v)\in \bar{M}}P_{u,v}^{\mathrm{min}}$.
\end{enumerate}
where here $P_{u,v}^{\mathrm{min}}\subseteq \mathcal{E}$ is a minimum-length path between $u$ and $v$ in $\mathcal{G}$.
See Theorem 12.10 of \cite{korte2011combinatorial} for a proof of this reduction, where their minimum-weight $T$-join is our MWEM, and their set $T$ corresponds to our $\mathcal{D}$.
See also \cite{barahona1982computational,berman1999tjoin} for alternative reductions and \cite{boyaci2022matchings} for a recent review.

Unfortunately, solving these three steps sequentially is quite computationally expensive; for example, just the cost of enumerating the edges in $\bar{\mathcal{G}}[\mathcal{D}]$ scales quadratically in the number of detection events $|\mathcal{D}|$, whereas we would ideally like a decoder with an expected running time that scales linearly in $|\mathcal{D}|$.
This sequential approach has nevertheless been widely used by QEC researchers, despite its performance being very far from optimal.

A significant improvement was introduced by Fowler~\cite{fowler2013minimum}.
A key observation made by Fowler was that, for QEC problems, typically only low-weight edges in $\bar{\mathcal{G}}[\mathcal{D}]$ are actually used by blossom.
Fowler's approach exploited this fact by setting an initial exploration radius in the detector graph, within which separate searches were used to construct some of the edges in $\bar{\mathcal{G}}[\mathcal{D}]$.
This exploration radius was then adaptively increased as required by the blossom algorithm.
Our approach, sparse blossom, is inspired by Fowler's work but is different in many of the details.
Before introducing sparse blossom, we will next give an overview of the standard blossom algorithm.

\subsection{The blossom algorithm}

The blossom algorithm, introduced by Jack Edmonds~\cite{edmonds1965paths,edmonds1965maximum}, is a polynomial-time algorithm for finding a minimum-weight perfect matching in a graph.
In this section we will outline some of the key concepts in the original blossom algorithm.
We will not explain the original blossom algorithm in full, since there is significant overlap with our sparse blossom algorithm, which we describe in \sect{sparse_blossom}.
While this section provides motivation for sparse blossom, it is not a prerequisite for the rest of the paper, so the reader may wish to skip straight to \sect{sparse_blossom}.
We refer the reader to references \cite{edmonds1965paths,edmonds1965maximum,galil1986efficient,kolmogorov2009blossom} for a more complete overview of the blossom algorithm.

We will first introduce some terminology.
Given some matching $M\subseteq E$ in a graph $G=(V, E)$, we say that an edge in $E$ is \textit{matched} if it is also in $M$, and unmatched otherwise, and a node is matched if it is incident to a matched edge, and unmatched otherwise.
A maximum cardinality matching is a matching that contains as many edges as possible.
An \textit{augmenting path} is a path $P\subseteq E$ which alternates between matched and unmatched edges, and begins and terminates at two distinct unmatched nodes.

Given an augmenting path $P$ in $G$, we can always increase the cardinality of the matching $M$ by one by replacing $M$ with the new matching $M^\prime=M\oplus P$.
We refer to this process, of adding each unmatched edge in $P$ to $M$ and removing each matched edge in $P$ from $M$, as \textit{augmenting} the augmenting path $P$ (see \fig{augmenting_paths_and_alternating_trees}(a)).
Berge's theorem states that a matching has maximum cardinality if and only if there is no augmenting path~\cite{berge1957two}.

\subsubsection{Solving the maximum cardinality matching problem}
\label{sec:max_cardinality_matching}

\begin{figure}
    \centering
    \includegraphics[width=0.9\columnwidth]{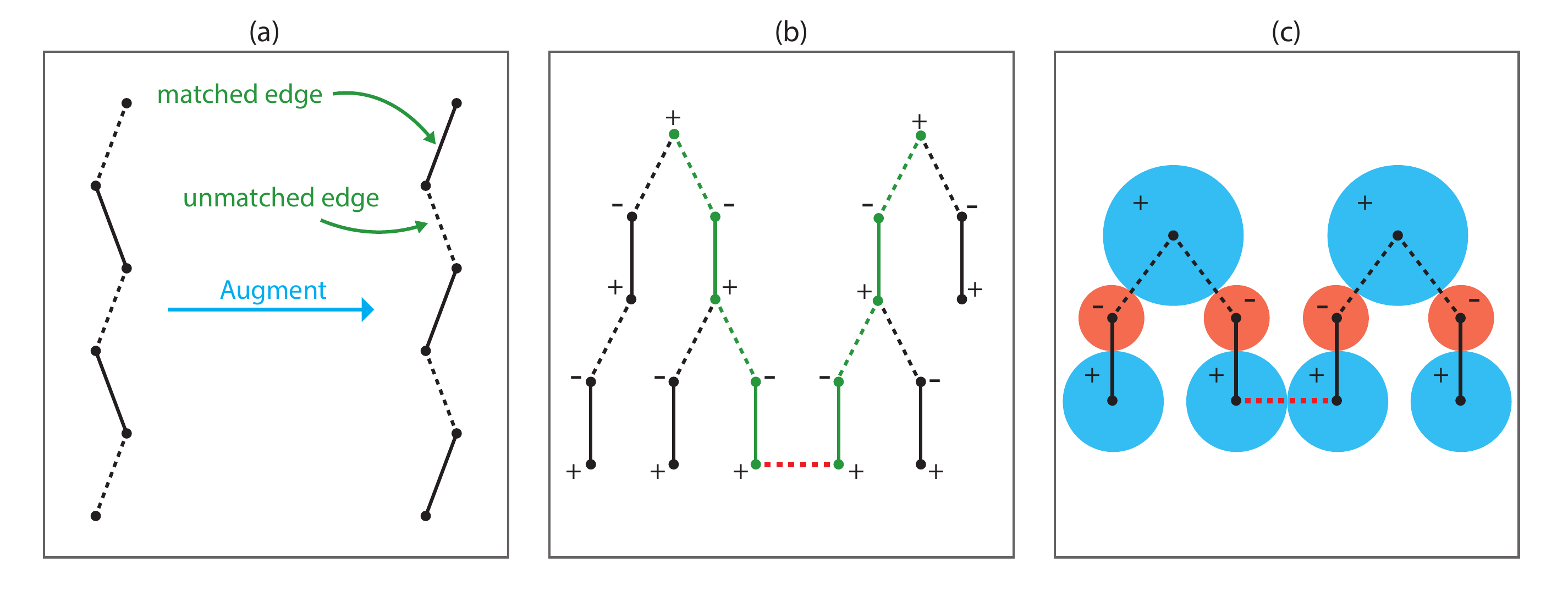}
    \caption{(a) Augmenting an augmenting path. Matched edges become unmatched, and unmatched edges become matched. (b) Examples of two alternating trees in the blossom algorithm for finding a maximum matching. Each tree has one unmatched node. The two trees have become connected via the red dashed edge. The path between the roots of the two trees, through the green edges and red edge, is an augmenting path. (c) An example of two alternating trees in the blossom algorithm for finding a minimum-weight perfect matching. Each node $v$ now has a dual variable $y_v$ which, when $y_v$ is positive, we can interpret as the radius of a region centred on the node. A new edge $(u, v)$ with weight $w_{u, v}$ can only be explored by the alternating tree if it is \textit{tight}, meaning that the dual variables $y_u$ and $y_v$ satisfy $y_u+y_v=w_{u, v}$.}
    \label{fig:augmenting_paths_and_alternating_trees}
\end{figure}

We will now give an overview of the original \textit{unweighted} version the blossom algorithm, which finds a maximum cardinality matching (as introduced by Edmonds in \cite{edmonds1965paths}).
The unweighted blossom algorithm is used as a subroutine by the more general blossom algorithm for finding a minimum-weight perfect matching (discovered, also by Edmonds, in \cite{edmonds1965maximum}), which we will outline in \sect{weighted_blossom}.
The algorithm is motivated by Berge's theorem.
Starting with a trivial matching, it proceeds by finding an augmenting path, augmenting the path, and then repeating this process until no augmenting path can be found, at which point we know that the matching is maximum.
Augmenting paths are found by constructing \textit{alternating trees} within the graph.
An alternating tree $T$ in the graph $G$ is a tree subgraph of $G$ with an unmatched node as its root, and for which every path from root to leaf alternates between unmatched and matched edges, see \fig{augmenting_paths_and_alternating_trees}(b).
There are two types of nodes in $T$: ``outer'' nodes (labelled ``$+$'') and ``inner'' nodes (labelled ``$-$'').
Each inner node is separated from the root node by an odd-length path, whereas each outer node is separated by an even-length path.
Each inner node has a single child (an outer node). 
Each outer node can have any number of children (all inner nodes).
All leaf nodes are outer nodes.

Initially, every unmatched node is a trivial alternating tree (a root node).
To find an augmenting path, the algorithm searches the neighboring nodes in $G$ of the outer nodes in each tree $T$.
If, during this search, an edge $(u, v)$ is found such that $u$ is an outer node of $T$ and $v$ is an outer node of some other tree $T^\prime \not = T$ then an augmenting path has been found, which connects the roots of $T$ and $T^\prime$, see \fig{augmenting_paths_and_alternating_trees}(b).
This path is augmented, the two trees are removed, and the search continues.
If an edge $(u, v)$ is found between an outer node $u$ in $T$ and a matched node $v$ not in any tree (i.e.~$v$ is matched), then $v$ and its match are added to $T$.
Finally, if an edge $(u, v)$ is found between two outer nodes of the same tree then an odd-length cycle has been found, and forms a \textit{blossom}.
A key insight of Edmonds was that a blossom can be treated as a virtual node, which can be matched or belong to an alternating tree like any other node.
However, we will explain how blossoms are handled in more detail in the context of our sparse blossom algorithm in \sect{sparse_blossom}.

\subsubsection{Solving the minimum-weight perfect matching problem}\label{sec:weighted_blossom}

The extension from finding a maximum cardinality matching to finding a minimum-weight perfect matching is motivated by formulating the problem as a linear program~\cite{edmonds1965maximum}.
Constraints are added on how the alternating trees are allowed to grow, and these constraints ensure that the weight of the perfect matching is minimal once it has been found.
The formulation of the problem as a linear program is not required either to understand the algorithm, or for the proof of correctness.
However, it does provide useful motivation, and the constraints and definitions used in the linear program are also used in the blossom algorithm itself.
We will therefore describe the linear program here for completeness.

We will denote the boundary edges of some subset of the nodes $S\subseteq V$ by $\delta(S)\coloneqq \{(u,v)\in E\ |\ u\in S, v\in V\setminus S\}$, and will let $\mathcal{O}$ be the set of all subsets of $V$ of odd cardinality at least three, i.e.~$\mathcal{O}\coloneqq \{o\subseteq V:|o|>1,|o|\mod 2 = 1\}$.
We denote the edges incident to a single node $v$ by $\delta(v)$ (i.e.~$\delta(v)=\delta(\{v\})$).
We will use an incidence vector $\mathbf{x}\in \{0,1\}^{|E|}$ to represent a matching $M\subseteq E$ where $x_e=1$ if $e\in M$ and $x_e=0$ if $e\notin M$.
We denote the weight of an edge $e\in E$ by $w_e$.
The minimum-weight perfect matching problem can then be formulated as the following \textit{integer} program:
\begin{subequations}
\begin{align}
    \text{Minimise}\quad &\sum_{e\in E}w_ex_e \\
    \text{subject to}\quad &\sum_{e\in\delta(v)}x_e=1\quad \forall v \in V \\
    & x_e \in \{0,1\} \quad\forall e \in E \label{eq:integrality}
\end{align}
\end{subequations}
Edmonds introduced the following linear programming relaxation of the above integer program:
\begin{subequations}
\begin{align}
    \text{Minimise}\quad &\sum_{e\in E}w_ex_e \\
    \text{subject to}\quad &\sum_{e\in\delta(v)}x_e=1\quad \forall v \in V \\
    &\sum_{e\in\delta(S)}x_e\geq 1\quad\forall S\in\mathcal{O}\label{eq:odd_constraints} \\ 
    & x_e \geq 0 \quad\forall e \in E \label{eq:variable_constraints}.
\end{align}
\end{subequations}
Note that the constraints in \eq{odd_constraints} are satisfied by \textit{any} perfect matching, but Edmonds showed that adding them ensures that the linear program has an integral optimal solution.
In other words, the integrality constraint (\eq{integrality}) can be replaced by the inequalities in \eq{odd_constraints} and \eq{variable_constraints}.

Every linear program (referred to as the \textit{primal} linear program, or primal problem) has a \textit{dual} linear program (or dual problem).
The dual of the above primal problem is:
\begin{subequations}
\label{eq:dual_problem}
\begin{align}
    \text{Maximise}\quad & \sum_{v\in V}y_v + \sum_{S\in\mathcal{O}}y_S \\
    \text{subject to}\quad & slack(e)\geq 0\quad \forall e \in E \label{eq:non_negative_slacks} \\
    &y_S\geq 0\quad \forall S \in \mathcal{O}\label{eq:non_negative_dual_variable}
\end{align}
\end{subequations}

where the \textit{slack} of an edge is defined as
\begin{equation}
    slack(e)\coloneqq w_e - \sum_{u\in e}y_u - \sum_{S\in \mathcal{O}:e\in\delta(S)}y_S\label{eq:slack}.
\end{equation}
We say that an edge is \textit{tight} if it has zero slack.
Here we have defined a dual variable $y_v\in\mathbb{R}$ for each node $v\in V$, as well as a dual variable $y_S\in\mathbb{R}$ for each set $S\in\mathcal{O}$.
While each variable $y_S$ is constrained to be non-negative (\eq{non_negative_dual_variable}), each $y_v$ is permitted to take any value.
Although we have an exponential number of $y_S$ variables, this turns out not to be an issue since only $O(|V|)$ are non-zero at any given stage of the blossom algorithm.

We now recall some terminology and general properties of linear programs (see \cite{matouvsek2007understanding,korte2011combinatorial} for more details).
A solution of a linear program is \textit{feasible} if it satisfies the constraints of the linear program.
Without loss of generality, we assume that the primal linear program is a minimisation problem (in which case its dual is a maximisation problem).
By the \textit{strong duality theorem}, if both the primal and the dual linear program have a feasible solution, then they both also have an \textit{optimal} solution. 
Furthermore, the minimum of the primal problem is equal to the maximum of its dual, providing a ``numerical'' proof of optimality.

We can obtain a ``combinatorial'' proof of optimality for any linear program using the \textit{complementary slackness} conditions.
Each constraint in the primal problem is associated with a variable of the dual problem (and vice versa).
Let us associate the $i$th primal constraint with the $i$th dual variable (and vice versa).
The complementary slackness conditions state that, if and only if we have a pair of optimal solutions, then if the $i$th dual variable is greater than zero then the $i$th primal constraint is satisfied with equality.
Similarly, if the $i$th primal variable is greater than zero then the $i$th dual constraint is satisfied with equality.
More concretely, for the specific primal-dual pair of linear programs we are considering, the complementary slackness conditions are:
\begin{align}
    slack(e)>0&\implies x_e=0 \label{eq:comp_slack_1} \\
    y_S>0 &\implies \sum_{e\in\delta(S)}x_e=1\quad\quad (S\in\mathcal{O}) \label{eq:comp_slack_2}
\end{align}
These conditions are used as a stopping rule in the blossom algorithm (with \eq{comp_slack_1} satisfied throughout) and provide a proof of optimality.

While it is convenient to use the strong duality theorem, since it applies to \textit{any} linear program, its correctness is not immediately intuitive and its proof is quite involved (see \cite{matouvsek2007understanding,korte2011combinatorial}).
Fortunately, we can obtain a simple proof of optimality of the minimum-weight perfect matching problem directly, without the need for the strong duality theorem~\cite{galil1986efficient}.
First, we note that for any feasible dual solution, we have that \textit{any} perfect matching $N$ satisfies
\begin{equation}
    \sum_{e\in N}w_e=\sum_{e\in N}\left(slack(e)+\sum_{v\in e}y_v + \sum_{S\in\mathcal{O}:e\in\delta(S)} y_S \right) \geq \sum_{v\in V}y_v + \sum_{S\in\mathcal{O}}y_S\label{eq:dual_lower_bound},
\end{equation}
where here the equality is from the definition of $slack(e)$ and the inequality uses \eq{non_negative_slacks} and \eq{non_negative_dual_variable} (i.e.~the fact that the dual solution is feasible) and the fact that $N$ is a perfect matching. 
However, if we have a perfect matching $M$ which additionally satisfies \eq{comp_slack_1} and \eq{comp_slack_2} we instead have
\begin{equation}
    \sum_{e\in M}w_e=\sum_{v\in V}y_v + \sum_{S\in\mathcal{O}}y_S,
\end{equation}
and thus the perfect matching $M$ has minimal weight.

So far in this section, we have only considered the case that each edge is a \textit{pair} of nodes (a set of cardinality two).
Let us now consider the more general case (required for decoding) where we can also have half-edges.
More specifically, we now have the edge set $E=E_1\cup E_2$ where each $(u,v)\in E_2$ is a regular edge and each $(u,)\in E_1$ is a half-edge (a node set of cardinality one).
We note that a perfect matching is now defined as a subset of this more general edge set, but its definition is otherwise unchanged (a perfect matching is an edge set $M\subseteq E\coloneqq E_1\cup E_2$ such that each node is incident to exactly one edge in $M$).
We extend our definition of $\delta(S)$ to be 
\begin{equation}
    \delta(S)\coloneqq \{(u,v)\in E_2\ |\ u\in S, v\in V\setminus S\}\cup \{(u,)\in E_1\ |\ u\in S\}.
\end{equation}
With this modification, the simple proof of correctness above still holds and the $slack(e)$ of a half-edge $e\in E_1$ is well defined by \eq{slack}.

The blossom algorithm for finding a minimum-weight perfect matching starts with an empty matching and a feasible dual solution, and iteratively increases the cardinality of the matching and the value of the dual objective while ensuring the dual problem constraints remain satisfied.
Eventually, we will have a pair of feasible solutions to the primal and dual problem satisfying the complementary slackness conditions (\eq{comp_slack_1} and \eq{comp_slack_2}) at which point we know we have a perfect matching of minimal weight. The algorithm proceeds in stages, where each stage consists of a ``primal update'' and a ``dual update''.
We repeat these primal and dual updates until no more progress can be made at which point, provided the graph admits a perfect matching, the complementary slackness conditions will be satisfied and so the minimum-weight perfect matching has been found.
We will now outline the primal and dual update in more detail.

In the \textit{primal update}, we consider only the subgraph $H$ of $G$ consisting of \textit{tight} edges and try to find a matching of higher cardinality, essentially by running a slight modification to the unweighted blossom algorithm on this subgraph.
In \cite{kolmogorov2009blossom}, the four allowed operations in the primal update are referred to as ``GROW'', ``AUGMENT'', ``SHRINK'' and ``EXPAND''.
The first three of these already occur in the unweighted variant of blossom discussed in \sect{max_cardinality_matching}.
The GROW operation consists of adding a matched pair of nodes to an alternating tree.
AUGMENT is the process of augmenting the path between the roots of two trees when they become connected.
SHRINK is the name for the process of forming a blossom when an odd length cycle is found.
The operation that differs slightly in the weighted variant is EXPAND.
This EXPAND operation can occur whenever the dual variable $y_S$ for a blossom $S$ becomes zero; when this happens the blossom is removed, the odd-length path through the blossom is added into the alternating tree, and nodes in the even-length path become matched to their neighbours.
This differs slightly from the unweighted variant as we described it, where blossoms are only expanded when a path they belong to becomes augmented (at which point \textit{all} the nodes in a blossom cycle become matched).
We refer the reader to \cite{kolmogorov2009blossom} for a more complete description of these operations in the primal update (and associated diagrams), although we reiterate that very similar concepts will be covered in more detail when we describe sparse blossom in \sect{sparse_blossom}.

In the \textit{dual update}, we try to increase the dual objective by updating the value of the dual variables, ensuring that edges in alternating trees and blossoms remain tight, and also ensuring that the dual variables remain a feasible solution to the dual problem (the inequalities \eq{non_negative_slacks} and \eq{non_negative_dual_variable} must remain satisfied).
Loosely speaking, the goal of the dual update is to increase the dual objective in such a way that more edges become tight, while ensuring existing alternating trees, blossoms and matched edges remain intact.
The only dual variables we update are those belonging to nodes in an alternating tree.
For each alternating tree $T$ we choose a dual change $\delta_T\geq 0$ and we \textit{increase} the dual variable of every outer node $u$ with $y_u\coloneqq y_u+\delta_T$ but \textit{decrease} the dual variable of every inner node $u$ with $y_u\coloneqq y_u-\delta_T$.
Recall that each node in $T$ is either a regular node or a blossom, and if the node is a blossom then we are changing the \textit{blossom}'s dual variable (while leaving the dual variables of the nodes it contains unchanged).
Note that this change ensures that all tight edges within a given alternating tree remain tight, but since outer node dual variables are increasing, it is possible that some of their neighbouring (non-tight) edges may \textit{become} tight (hopefully allowing us to find an augmenting path between alternating trees in the next primal update).
The constraints of the dual problem (the inequalities \eq{non_negative_slacks} and \eq{non_negative_dual_variable}) impose constraints on the choice of $\delta_T$; in particular, the slacks of all edges must remain non-negative, and \textit{blossom} dual variables must also remain non-negative.

There are many different valid strategies that can be taken for the dual update.
In a \textit{single tree} approach, we pick a single tree $T$ and update the dual variables only of the nodes in $T$ by the maximum amount $\delta_T$ such that the constraints of the dual problem remain satisfied (e.g.~we change the dual variables until an edge becomes tight or a blossom dual variable becomes zero).
In a \textit{multiple tree fixed $\delta$} approach, we update the dual variables of \textit{all} alternating trees by the same amount $\delta_T$ (again by the maximum amount that ensures the dual constraints remain satisfied).
In a \textit{multiple tree variable $\delta$} approach, we choose a different $\delta_T$ for each tree $T$.
Our variant of the blossom algorithm (sparse blossom) uses a multiple tree fixed $\delta$ approach.
This is a key difference with Refs.~\cite{fowler2012towards, fowler2012timing_analysis, fowler2013minimum}, which instead use a single tree approach.
See \cite{kolmogorov2009blossom} for a more detailed discussion and comparison of these different strategies.

In \fig{augmenting_paths_and_alternating_trees}(c) we give an example with two alternating trees, and visualise a dual variable as the radius of a circular region centred on its node.
Visualising dual variables this way, an edge between two trivial nodes is tight if the regions at its endpoints touch.
In this example, we update the dual variables (radiuses) until the two alternating trees touch, at which point the edge joining the two trees becomes tight, and we can augment the path between the roots of the two trees.
Note that we can only visualise dual variables as region radiuses like this when they are non-negative.
While dual variables of blossoms are always non-negative (as imposed by \eq{non_negative_dual_variable}), dual variables of regular nodes \textit{can} become negative in general.
However, when running the blossom algorithm on a \textit{path graph}, the dual variable of every regular node is also always non-negative, owing to the structure of the graph.
This can be understood as follows.
Consider any regular inner node $v$ that is not a blossom, which by definition must have exactly one child outer node $w$ in its alternating tree (its match), as well as its one parent outer node $u$.
Recall that the path graph is a complete graph where the weight $w(x,y)$ of each edge $(x,y)$ is the length of shortest path between nodes $x$ and $y$ in some other graph (e.g.~in our case always the detector graph).
Therefore there is also an edge $(u,w)$ in the path graph with weight $w(u,w)\leq w(u,v)+w(v,w)$, since we know that there is at least one path from $u$ to $w$ of length $w(u,v)+w(v,w)$, corresponding to the union of the shortest path from $u$ to $v$ and the shortest path from $v$ to $w$.
Therefore, we cannot have $y_v<0$ without having $slack((u,w))<0$ which would violate \eq{non_negative_slacks}.
More specifically, if $y_v=0$ then we know that the edge $(u,w)$ \textit{must} be tight, which means we can form a new blossom from the blossom cycle $(u,v,w)$ and this blossom can become an \textit{outer} node in the (possibly now trivial) alternating tree.

\section{Sparse Blossom}\label{sec:sparse_blossom}

The variant of the blossom algorithm we introduce, which we call sparse blossom, directly solves the minimum-weight embedded matching problem relevant to quantum error correction.
Sparse blossom does not have a separate Dijkstra step for constructing edges in the path graph $\bar{\mathcal{G}}[\mathcal{D}]$.
Instead, shortest path information is recovered as part of the growth of alternating trees themselves.
Put another way, we only discover and store an edge $e\in\bar{\mathcal{E}}$ in $\bar{\mathcal{G}}[\mathcal{D}]$ exactly if and when it is needed by the blossom algorithm; the edges that we track at any point in the algorithm correspond exactly to the subset of \textit{tight} edges in $\bar{\mathcal{E}}$ being used to represent an alternating tree, a blossom or a match.
This leads to very large speedups relative to the sequential approach, where all edges in $\bar{\mathcal{E}}$ are found using Dijkstra searches, despite the vast majority never becoming tight edges in the blossom algorithm.
We name the algorithm \textit{sparse} blossom, since it exploits the fact that only a small fraction of the detector nodes correspond to detection events for typical QEC problems (and detection events can be paired up locally), and for these problems our approach only ever inspects a small subset of the nodes and edges in the detector graph.

Before explaining sparse blossom and our implementation, we will first introduce and define some concepts.

\subsection{Key concepts}

\subsubsection{Graph fill regions}

A \textit{graph fill region} $R$ of radius $y_R$ is an exploratory region of the detector graph.
A graph fill region $R$ contains the nodes in the detector graph which are within distance $y_R$ of its \textit{source}.
The source of a graph fill region is either a single detection event, or the surface of other graph fill regions forming a \textit{blossom}.
We will define blossoms later on, however for the case that the graph fill region $R$ has a single detection event $u$ as its source, every node that is within distance $y_R$ of $u$ is contained in $R$.
Recall that the distance $D(u,v)$ between two nodes $u$ and $v$ is the sum of the weights of edges along the weighted shortest path between them.
Note that a graph fill region in sparse blossom is analogous to a node or blossom in the standard blossom algorithm, and the graph fill region's radius is analogous to a node or blossom's dual variable in standard blossom.
The sparse blossom algorithm proceeds along a timeline (see \sect{timeline}), and the radius of each graph fill region can have one of three growth rates: +1 (growing), -1 (shrinking) or 0 (frozen).
Therefore, at any time $t$ we can represent the radius of a region using an equation $y_R=mt+c$, where $m\in\{-1, 0, 1\}$.
We will at times refer to a graph fill region just as a \textit{region} when it is clear from context.

We denote by $\mathcal{D}(R)$ the set of detection events in a region $R$.
When a region contains only a single detection event, $|\mathcal{D}(R)|=1$, we refer to it as a trivial region.
A region can contain multiple detection events if it has a blossom as its source.
As well as its radius equation, each graph fill region may also have blossom children and a blossom parent (both defined in \Cref{sec:blossoms}).
It also has a \textit{shell area}, stored as a stack.
The shell area of a region is the set of detector nodes it contains, excluding the detector nodes contained in its blossom children.
We say that a graph fill region is \textit{active} if it does not have a blossom parent.
We will let $\mathcal{R}$ denote the set of all graph fill regions.

\begin{figure}
    \centering
    \includegraphics[width=\columnwidth]{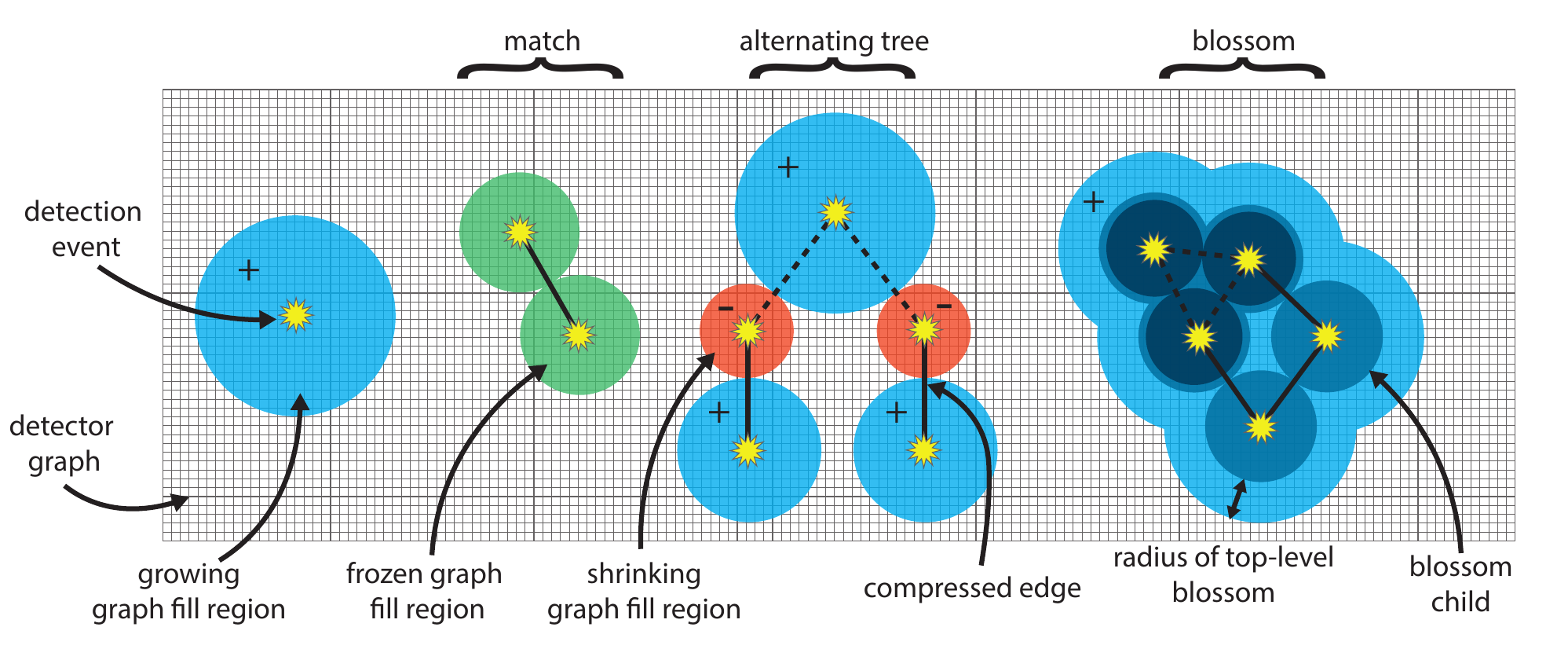}
    \caption{Key concepts in sparse blossom}
    \label{fig:sparse_blossom_concepts}
\end{figure}

\subsubsection{Compressed edges}

A \textit{compressed edge} represents a path through the detector graph between two detection events, or between a detection event and the boundary.
Given a path $P_{u,v}\subseteq \mathcal{E}$ between $u$ and $v$, where $u$ is a detection event and $v$ is either a detection event or denotes the boundary, the compressed edge $\beta(P_{u,v})$ associated with $P_{u,v}$ is the pair of nodes $(u, v)$ at the endpoints of $P_{u,v}$, as well as the set of logical observables $l(P_{u,v})\coloneqq \bigoplus_{e\in P_{u,v}}l(e)$ flipped by flipping all edges in $P_{u,v}$.
The compressed edge $\beta(P_{u,v})$ is therefore a compressed representation of the path $P_{u,v}$ containing all the information relevant for error correction and, for a given detector graph, can be stored using a constant amount of data (independent of the path length).
When the choice of path $P_{u,v}\subseteq \mathcal{E}$ for some given pair of detection events $(u,v)$ is clear from context, we may denote the compressed edge $\beta(P_{u,v})$ instead by $(u,v)$, and may also denote the set of logical observables $l(P_{u,v})$ by $l(u,v)$.
We define the \textit{length} of a compressed edge to be the distance $D(u,v)$ between its endpoints $u$ and $v$.

Every compressed edge $\beta(P_{u,v})$ in sparse blossom corresponds to a \textit{shortest} path $P_{u,v}$ between $u$ and $v$.
However, we can use a compressed edge to represent any path between $u$ and $v$, and when used in union-find (see \app{compressed_uf}) the path $P_{u,v}$ need not be minimum weight.
Compressed edges are used in the representation of several data structures in sparse blossom (alternating trees, matches and blossoms).
In particular, each compressed edge $(u,v)$ corresponds to an edge in the path graph $\bar{\mathcal{G}}[\mathcal{D}]$, but unlike in the standard serial approach to implementing the MWPM decoder, we only every discover and construct the small subset of the edges in $\bar{\mathcal{G}}[\mathcal{D}]$ needed by sparse blossom.

We will let $\Gamma$ be the set of all possible compressed edges that can represent shortest paths between detector nodes (i.e.~the set $\Gamma$ contains a compressed edge $(u,v)$ for each edge in $\bar{\mathcal{G}}[\mathcal{D}]$).
For a region $R$, we denote by $\delta_{\beta}(R)\subseteq \Gamma$ the \textit{boundary-compressed-edges} of $R$, defined as 
\begin{equation}
    \delta_{\beta}(R)\coloneqq \{(u,v)\in \Gamma\ |\ u \in \mathcal{D}(R), v \notin \mathcal{D}(R)\}.
\end{equation}

 \subsubsection{Region edges}

 A \textit{region edge} describes a relationship between two graph fill regions, or between a region and the boundary.
 We use $(A,B)$ to denote a region edge, where here $A$ and $B$ are both regions.
 Whenever we describe an edge between two regions, or between a region and the boundary, it is implied that it is a region edge.
 A region edge $(A, B)$ comprises its endpoints $A$ and $B$ as well as a compressed edge $\beta(P_{u,v})$ representing the shortest path between any detection event in $A$ and any detection event in $B$.
 We sometimes use the notation $(A_u,B_v)$ for a region edge to explicitly specify the two regions $A$ and $B$ along with the endpoints $(u,v)$ of the compressed edge $\beta(P_{u,v})$ associated with it.

 For any region edge that arises in sparse blossom, the following invariants always hold:
 \begin{enumerate}
     \item If $A$ and $B$ are both regions, then either $A$ and $B$ are both active regions (with no blossom-parent), or both have the same blossom-parent.
     \item The compressed edge $(u,v)$ associated with a region edge $(A_u,B_v)$ is always \textit{tight} (it would correspond to a tight edge in $\bar{\mathcal{G}}[\mathcal{D}]$ in blossom). More precisely, a compressed edge $(u,v)$ is tight if it satisfies 
     \begin{equation}
         D(u,v)=\sum_{R\in\mathcal{R}\ :\ (u,v)\in\delta_{\beta}(R)}y_R.
     \end{equation}
     In other words, if there is a region edge $(A,B)$, then regions $A$ and $B$ must be touching.
 \end{enumerate}

\subsubsection{Matches}

We call a pair of regions $A$ and $B$ that are matched to each other a \textit{match}.
If $A$ and $B$ are matched, this corresponds to the compressed edge between $A$ and $B$ being matched on the path graph.
The matched regions $A$ and $B$ must be touching, assigned a growth rate of zero (they are frozen), and must be joined by a region edge $(A,B)$ which we refer to as a \textit{match edge}.
An example of a match is shown in the middle of \fig{sparse_blossom_concepts}.
Initially, all regions are unmatched, and once the algorithm terminates, every region (either a trivial region or a blossom) is matched either to another region or to the boundary.

\subsubsection{Alternating trees}

An alternating tree is a tree where each node corresponds to an active graph fill region and each edge corresponds to a region edge.
We refer to each region edge in the alternating tree as a \textit{tree-edge}.
Two regions connected by a tree-edge must always be touching (since every tree-edge is a region edge).

An alternating tree contains at least one growing region and can also contain shrinking regions, and always contains exactly one more growing region than shrinking region.
Each growing region can have any number of children (each a \textit{tree-child}), all of which must be shrinking regions, and can have a single parent (a \textit{tree-parent}), also a shrinking region.
Each shrinking region has a single child, a growing region, as well as a single parent, also a growing region.
The leaves of an alternating tree are therefore always growing regions.
An example of an alternating tree is shown in \fig{sparse_blossom_concepts}.

\subsubsection{Blossoms}\label{sec:blossoms}

When two growing regions from within the same alternating tree hit each other they form a \textit{blossom}, which is a region containing an odd-length cycle of regions called a \textit{blossom cycle}.
More concretely, we will denote a blossom cycle as an ordered tuple of regions $(R_0,R_1,\ldots,R_{k-1})$ for some odd $k$ and each region $R_i$ in the blossom cycle is connected to each of its two neighbours by a region edge.
In other words, the blossom cycle has region edges $\{(R_i,R_{(i+1)\mod k})|i\in \{0,1,\ldots,k-1\}\}$.
An example of a blossom is shown on the right side of \fig{sparse_blossom_concepts}.

The regions in the blossom cycle are called the blossom's \textit{blossom-children}.
If a blossom $B$ has region $b$ as one of its blossom-children, then we say that $B$ is the \textit{blossom-parent} of $b$.
Neighbouring regions in a blossom cycle must be touching (as required by the fact that they are connected by a region edge).
Blossoms can also be nested; each blossom-child can itself be a blossom, with its own blossom-children.
For example, the top-left blossom-child of the blossom in \fig{sparse_blossom_concepts} is itself a blossom, with three blossom-children.
A blossom \textit{descendant} of a blossom $B$ is a region that is either a blossom child of $B$ or is recursively a descendant of any blossom child of $B$.
Similarly, a blossom \textit{ancestor} of $B$ is a region that is either the blossom parent of $B$ or (recursively) an ancestor of the blossom parent of $B$.
The radius $y_B$ of a blossom $B$ (its dual variable in the standard blossom algorithm) is the distance it has grown since it formed (the minimum distance between a point on its surface and any point on its source).
This is visualised as the distance across the shell of the blossom in \fig{sparse_blossom_concepts}.

We say that a detector node $u$ is \textit{contained} in a region $R$ if $u$ is in the shell area of $R$.
A detector node can only be contained in at most one region (shell areas are disjoint).
If a detector node is not contained in a region we say that the node is \textit{empty}, and otherwise it is \textit{occupied}.
We say that a detector node $u$ is \textit{owned} by a region $R$ either if $u$ is contained in $R$, or if $u$ is contained in a blossom descendant of $R$.
The distance $D_S(R, u)$ between a detector node and the surface of a region $R$ is 
\begin{equation}
    D_S(R, u)=\min_{x\in \mathcal{D}(R)}\left(D(u,x)-\sum_{A\in\mathcal{R}:x\in\mathcal{D}(A), A\leq R}y_A\right)
\end{equation}
where here $A\leq R$ denotes that $A$ is either a descendant of $R$ or $A=R$.
If $D_S(R,u)<0$ then detector node $u$ is owned by region $R$, if $D_S(R,u)>0$ then detector node $u$ is not owned by $R$, whereas if $D_S(R,u)=0$ then $u$ may or may not be owned by $R$ (it is on the surface of $R$).

\subsubsection{The timeline}\label{sec:timeline}

The algorithm proceeds along a timeline with the time $t$ increasing monotonically as different events occur and are processed.
Examples of events include a region arriving at a node, or colliding with another region.
The time that each event occurs is determined based on the radius equations of the regions involved, as well as the structure of the graph.
The algorithm terminates when there are no more events left to be processed, which happens when all regions have been matched, and have therefore become frozen.

\subsection{Architecture}

Sparse blossom is split into different components: a \textit{matcher}, a \textit{flooder} and a \textit{tracker}.
Each component has different responsibilities.
The matcher is responsible for managing the structure of the alternating trees and blossoms, without knowledge of the underlying structure of the detector graph.
The flooder handles how graph fill regions grow and shrink in the detector graph, as well as noticing when a region collides with another region or the boundary.
When the flooder notices a collision involving a region, or when a region reaches zero radius, the flooder notifies the matcher, which is then responsible for modifying the structure of the alternating trees or blossoms.
The tracker is responsible for handling \textit{when} the different events occur, and ensures that the flooder handles events in the correct order, with the help of a single priority queue.
A priority queue is a type of queue, holding a collection of items, with the property that an item removed from the queue (with the ``extract-min'' operation) is guaranteed to have the highest priority among all items in the queue (in our case each item is an event, with earlier event times corresponding to a higher priority).
The tracker uses this priority queue to inform the flooder when every event should be handled.

\subsection{The matcher}

At the initialisation stage of the algorithm, every detection event is initialised as the source of a growing region, a trivial alternating tree.
As these regions grow and explore the graph, they can hit other (growing or frozen) regions, as well as the boundary, until eventually all regions are matched and the algorithm terminates.
A growing region cannot hit a shrinking region, since shrinking regions recede exactly as quickly as growing regions expand.

When the flooder notices that a growing region $R$ has hit another (growing or frozen) region $R^\prime$ or the boundary, it finds the \textit{collision edge} and gives it to the matcher.
The collision edge is a region edge between $R$ and $R^\prime$ (or, if $R$ hit the boundary, then between $R$ and the boundary).
The collision edge can be constructed by the flooder from local information at the point of collision, as will be explained in \sect{flooder}, and it is used by the matcher when handling events that change the structure of the alternating tree (which we refer to as \textit{alternating tree events}).
The matcher is responsible for handling alternating tree events, as well as for recovering the pairs of matched detection events once all regions have been matched.

\subsubsection{Alternating tree events}\label{sec:alt_tree_events}

\begin{figure}
    \centering
    \includegraphics[width=\columnwidth]{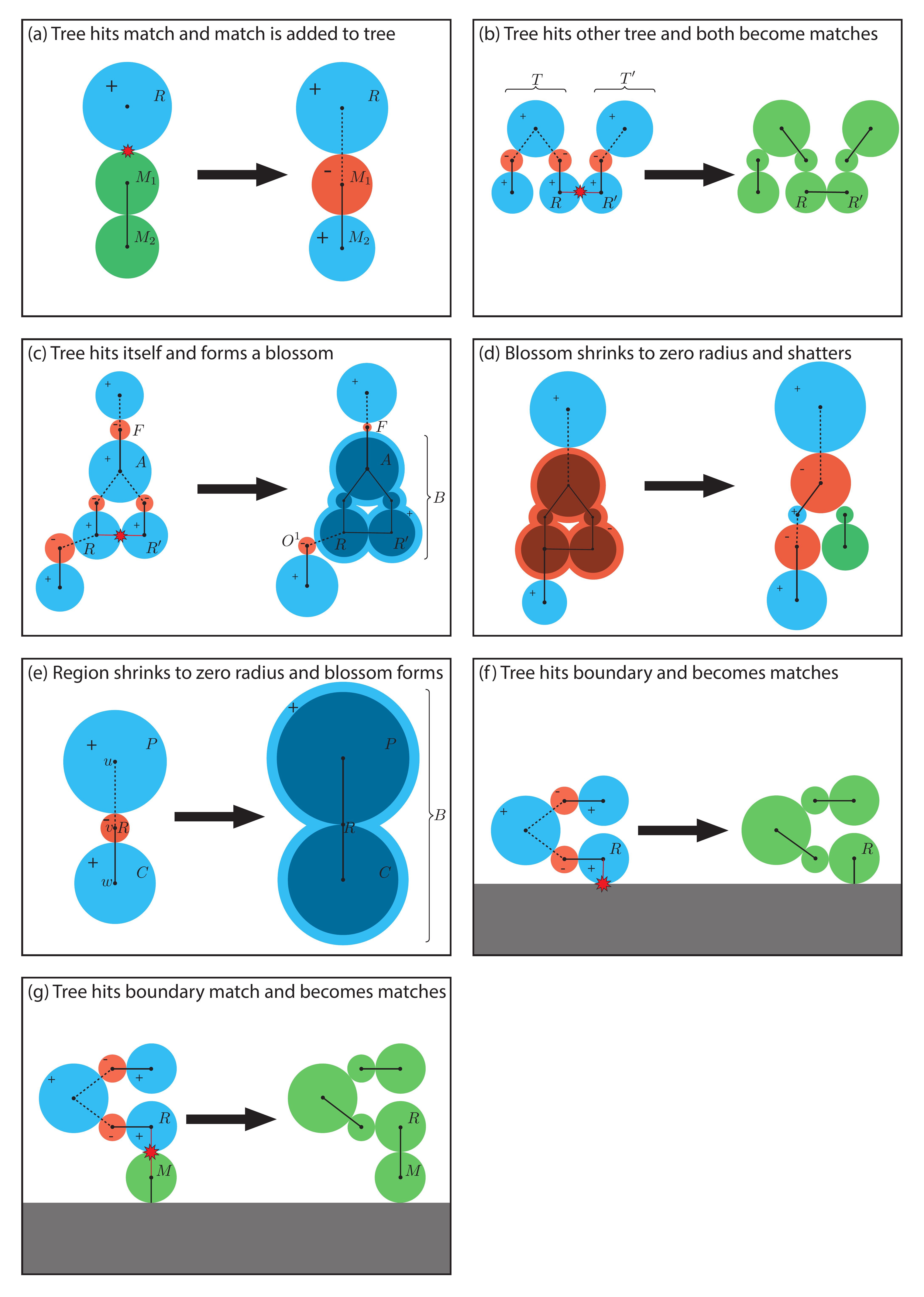}
    \caption{The main events that change the structure of alternating trees. For clarity, the background detector graph has been omitted.
    Each node corresponds to a detection event, and each edge corresponds to a compressed edge. Some labels (e.g.~of regions) have been included in the diagrams and correspond to those referred to in the main text in \sect{alt_tree_events}.}
    \label{fig:matcher_events}
\end{figure}

There are seven different types of events that can change the structure of an alternating tree, and which are handled by the matcher, shown in \fig{matcher_events}:
\begin{enumerate}[label=(\alph*)]
\item A growing region $R$ in an alternating tree $T$ can hit a region $M_1$ that is matched to another region $M_2$. In this case, $M_1$ becomes a tree-child of $R$ in $T$ (and starts shrinking), and $M_2$ becomes a tree-child of $M_1$ in $T$ (and starts growing). The collision edge $(R, M_1)$ and match edge $(M_1, M_2)$ both become tree-edges ($R$ is the tree-parent of $M_1$, and $M_1$ is the tree-parent of $M_2$).
\item A growing region $R$ in an alternating tree $T$ hits a growing region $R^\prime$ in a different alternating tree $T^\prime$. When this happens, $R$ is matched to $R^\prime$ and the remaining regions in $T$ and $T^\prime$ also become matched. The collision edge $(R, R^\prime)$ becomes a match edge, and a subset of the tree-edges also become match edges. All the regions in $T$ and $T^\prime$ become frozen.
\item A growing region $R$ in an alternating tree $T$ can hit another growing region $R^\prime$ in the \textit{same} alternating tree $T$. This leads to an odd-length cycle of regions which form the blossom cycle $C$ of a new blossom $B$. The region edges (blossom edges) in the blossom cycle are formed from the collision edge $(R, R^\prime)$, as well as the tree-edges along the paths from $R$ and $R^\prime$ to their most recent common ancestor $A$ in $T$. The newly formed blossom becomes a growing node in $T$. When forming the cycle $C$, we define the \textit{orphans} $O$ to be the set of shrinking regions in $T$ but not $C$ that were each a child of a growing region in $C$. The orphans become tree-children of $B$ in $T$. The compressed edge associated with the new tree-edge $(B, F)$ (connecting the new blossom region to its tree-parent $F$) is just the compressed edge that was associated with the old tree-edge $(A,F)$. Similarly, the compressed edges connecting each orphan to its alternating tree parent remains unchanged (even though its parent \textit{region} becomes $B$).
In other words, if an orphan $O^i$ had been connected to its tree-parent $R^i$ by the tree-edge $(O^i_u, R^i_v)$ before the blossom formed, the new tree-edge connecting it to its new tree-parent $B$ will be $(O^i_u, B_v)$ once the blossom $B$ forms and the region $R^i$ becomes part of the blossom cycle $C$.
\item When a blossom $B$ in an alternating tree $T$ shrinks to a radius of zero, instead of the radius becoming negative the blossom must \textit{shatter}. 
When the blossom shatters, the odd-length path through its blossom cycle from the tree-child of $B$ to the tree-parent of $B$ is added to $T$ as growing and shrinking regions. The even length path becomes matches.
The blossom-edges in the odd length path become tree-edges, and some of the blossom-edges in the even length path become match edges (the remaining blossom-edges are forgotten).
Note that the endpoints of the compressed edges associated with the tree-edges joining $B$ to its tree-parent and tree-child are used to determine where and how the blossom cycle is cut into two paths.
\item When a trivial region $R$ shrinks to a radius of zero, instead of the radius becoming negative a blossom forms.
If $R$ has a child $C$ and parent $P$ in the alternating tree $T$, when $R$ has zero radius it must be that $C$ is touching $P$ (it is as if $C$ has collided with $P$). The newly formed blossom $B$ has the blossom cycle $(P,R,C)$. 
The old tree-edges $(P_u,R_v)$ and $(R_v,C_w)$ become blossom edges in the blossom cycle.
The blossom edge connecting $C$ with $P$ in the blossom cycle is computed from edges $(P_u,R_v)$ and $(R_v,C_w)$ and is $(C_w,P_u)$.
In other words, its compressed edge has endpoints $(w, u)$ with logical observables $l(u,v)\oplus l(v,w)$.
\item When a growing region $R$ in an alternating tree $T$ hits the boundary, $R$ matches to the boundary and the collision edge becomes the match edge. The remaining regions in $T$ also become matches.
\item When a growing region $R$ in an alternating tree $T$ hits a region $M$ that is matched to the boundary, then $M$ instead becomes matched to $R$ (and the collision edge becomes the match edge), and the remaining edges in $T$ also become matches.
\end{enumerate}

Some of these events involve changing the growth rate of regions (for example, two growing regions both become frozen regions when they match to each other).
Therefore, when handling each alternating tree event, the matcher informs the flooder of any required changes to region growth.

\subsubsection{Matched detection events from matched regions}

\begin{figure}
    \centering
    \includegraphics[width=0.8\columnwidth]{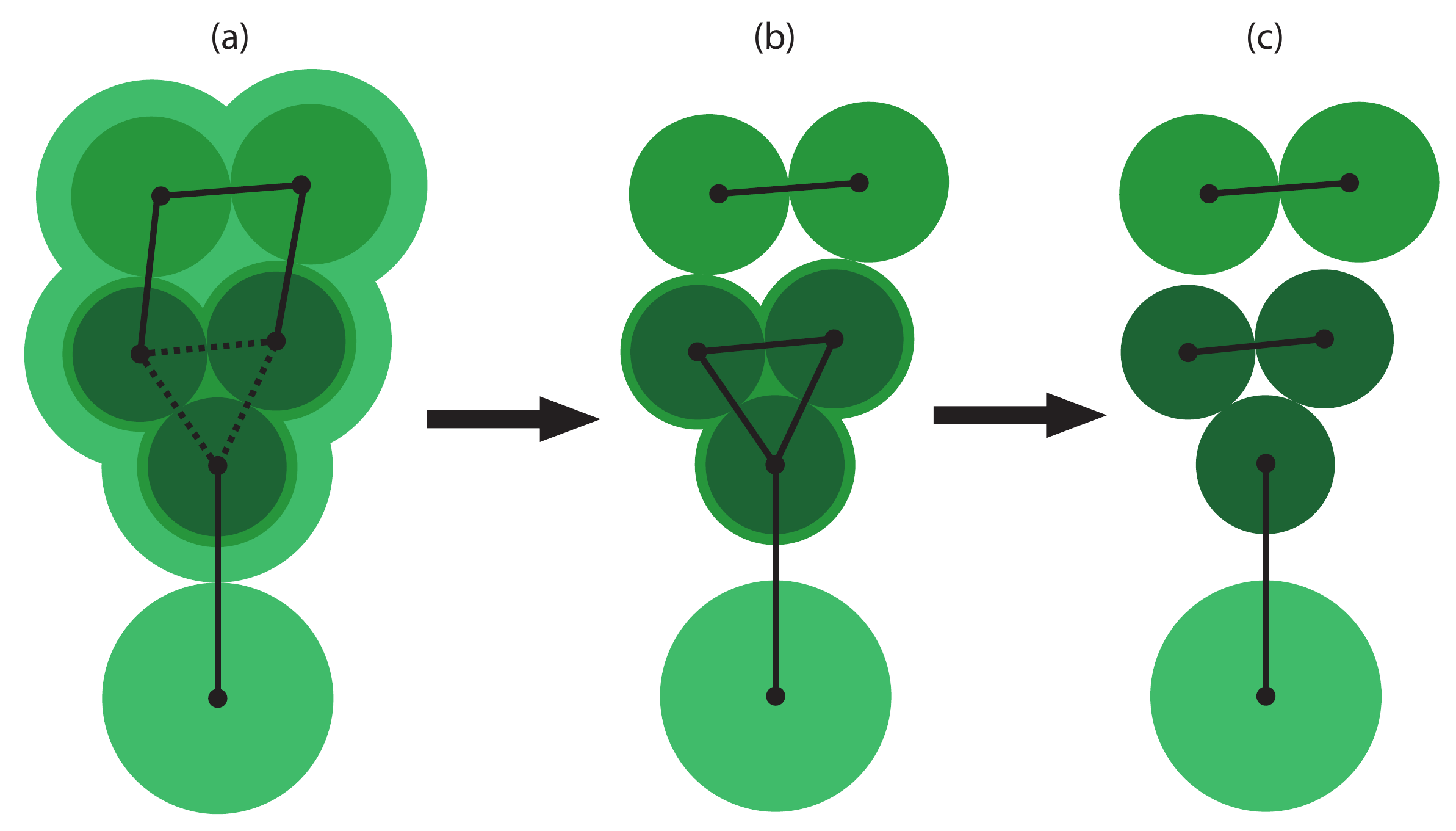}
    \caption{Shattering a matched blossom. Solid lines within a blossom are edges in the topmost blossom cycle. Dashed lines are edges in the blossom cycle of the blossom-child of the topmost blossom.}
    \label{fig:shatter_and_match}
\end{figure}

Provided there is a valid solution, eventually all regions become matched to other regions, or to the boundary.
However, some of these matched regions may be blossoms, not trivial regions.
To extract the compressed edge representing the match for each detection event instead, it is necessary to \textit{shatter} each remaining blossom, and match its blossom children, as shown in in \fig{shatter_and_match}.
Suppose a blossom $B$, with blossom cycle $C$, is matched to some other region $R$ with the match edge $(B_u, R_v)$, where we recall that $u$ is a detection event in $B$ and $v$ is a detection event in $R$.
We find the blossom child $c\in C$ of $B$ which contains the detection event $u$.
We shatter $B$ and match $c$ to $R$ with the compressed edge $(u, v)$.
The remaining regions in the blossom cycle $C$ are then split into neighbouring pairs, which become matches.
This process is repeated recursively until all matched regions are trivial regions.

\subsection{The flooder}\label{sec:flooder}

The flooder is responsible for managing how graph fill regions grow, shrink or collide in the detector graph, and is not concerned with the structure of the alternating trees and blossoms, which is instead handled by the matcher.
We refer to the events handled by the flooder as \textit{flooder events}.

Broadly speaking, we can categorise flooder events into four different types:
\begin{enumerate}
\item \label{grow_into_node} ARRIVE: A growing region $R$ can grow into an empty detector node $u$.
\item \label{leave_node} LEAVE: A shrinking region $R$ can leave a detector node $u$.
\item \label{collision} COLLIDE: A growing region can hit another region, or the boundary.
\item \label{implode} IMPLODE: A shrinking region can reach a radius of zero.
\end{enumerate}

Let us first consider what happens for ARRIVE and LEAVE events.
Neither of these types of events can change the structure of the alternating trees or blossoms, so the matcher does not need to be notified.
Instead, it is the flooder's responsibility to ensure that any new flooder events get scheduled (inserted into the tracker's priority queue) after the events have been processed.
When a region grows into a node $u$, the flooder \textit{reschedules} the node $u$, by notifying the tracker of the next flooder event that can occur along an edge adjacent to $u$ (either an ARRIVE or COLLIDE event, see \sect{reschedule_node}).
When a shrinking region leaves a node, the flooder immediately checks the top of the shell area stack and schedules the next LEAVE or IMPLODE event (see \sect{reschedule_shrink}).

The flooder only needs to notify the matcher of a COLLIDE or IMPLODE event, and when a collision occurs the flooder passes the collision edge to the matcher as well.
When either of these types of events occur, the matcher may change the growth rate of some regions when updating the structure of alternating trees or blossoms.
The matcher then notifies the flooder of any change of growth rate, which may require the flooder to reschedule some flooder events.
For example, if a region $R$ was shrinking, but then becomes frozen or starts growing, the flooder reschedules all nodes contained in $R$ (including blossom children, and their children etc.), to check for new ARRIVE or COLLIDE events.
When a region starts shrinking, then the flooder informs the tracker of the next LEAVE or IMPLODE event by checking the top of the shell area stack.

The correct ordering of these flooder events is ensured by the tracker, and we create a growing region for each detection event simultaneously at time $t=0$.
Our implementation therefore uses an alternating tree growth strategy that is analogous to what's described as a ``multiple tree approach with fixed $\delta$'' in \cite{kolmogorov2009blossom}.

\subsubsection{Rescheduling a node}\label{sec:reschedule_node}

\begin{figure}
    \centering
    \includegraphics[width=0.5\columnwidth]{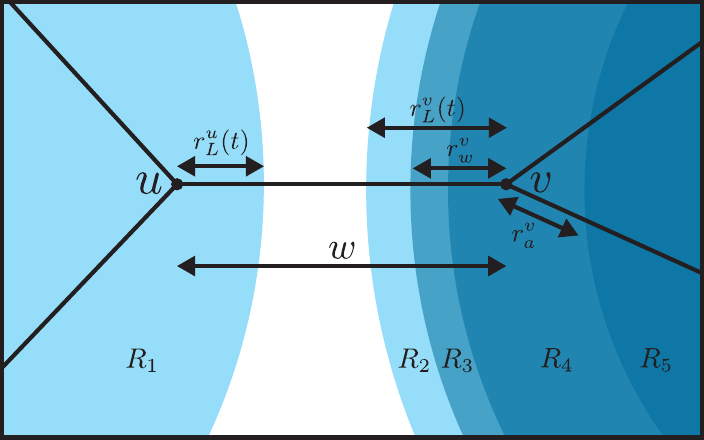}
    \caption{Two regions $R_1$ and $R_2$ colliding along an edge $(u, v)$. Node $u$ is contained in region $R_1$, which is an active region (no blossom parent). Node $v$ is contained in region $R_4$, and is owned by $R_4$ as well as its blossom ancestors $R_2$ and $R_3$. Region $R_4$ also has $R_5$ as one of its blossom children. We have labelled the local radius $r_L^u(t)$ of node $u$ and the local radius $r_L^v(t)$ of node $v$, as well as the wrapped radius $r_w^v$ of $v$ and the radius of arrival for $v$. The edge weight $w$ of the edge $(u, v)$ is also shown.}
    \label{fig:collision_along_edge}
\end{figure}

When the flooder \textit{reschedules} a node, it looks for an ARRIVE or COLLIDE event along each neighboring edge.
There will be an ARRIVE event along an edge if one node is occupied by a growing region and the other is empty.
There will be a COLLIDE event if both nodes are owned by active regions $(R_1, R_2)$ with growth rates $(1, 1)$, $(0, 1)$ or $(1, 0)$, or if one region is growing towards a boundary, for a half-edge.

In order to calculate \textit{when} an ARRIVE or COLLIDE event will occur along an edge, we use the \textit{local radius} of each node.
The \textit{local radius} $r_L^v(t)$ of an node $v$ is the amount that regions owning $v$ have grown beyond $v$ (see \fig{collision_along_edge}).
To define the local radius more precisely, we will need some more definitions.
The \textit{radius of arrival} $r_a^v$ for an occupied node $v$ contained in a region $R$ is the radius that $R$ had when it arrived at $v$.
We denote the radius of a region $A$ by $y_A(t)$ and we let $\mathcal{O}(v)$ be the set of regions that own a detector node $v$ (the region that $v$ is contained in, as well as its blossom ancestors).
The local radius is then defined as
\begin{equation}
    r_L^v(t)=-r_a^v + \sum_{A\in\mathcal{O}(v)}y_A(t).
\end{equation}
This definition can be understood by considering the example in \fig{collision_along_edge}, for which we recall that the radius of a blossom region can be visualized as the thickness of (i.e.~distance across) the shell of the blossom.
Both the local radius and radius of arrival of a node $v$ are defined to be zero if $v$ is empty.
Therefore, for an edge $(u, v)$ with weight $w$, the time of an ARRIVE or COLLIDE event can be found by solving $r_L^u(t)+r_L^v(t)=w$ for $t$.
The only situation in which this involves division is when the local radius of both nodes are growing (have gradient one), in which case the collision occurs at time $t_{\mathrm{collide}}=(w-r_L^u(0)-r_L^v(0))/2$.
However, provided all edges are assigned even integer weights all flooder events, including these collisions between growing regions, occur at integer times.

\subsubsection{Rescheduling a shrinking region}\label{sec:reschedule_shrink}

When a region is shrinking, we find the time of the next LEAVE or IMPLODE event by inspecting the shell area stack of the region.
If the stack is empty, or if the region has no blossom children and only a single node remains on the stack (the region's source detection event), then the next event is an IMPLODE event, the time of which can be found from the $x$-intercept of the region's radius equation.
Otherwise, the next event is a LEAVE event, with the node $u$ at the top of the stack leaving the region.
We find the time of this next LEAVE event using the local radius of $u$, by solving $r_L^u(t)=0$ for $t$.
Using this approach, shrinking a region is much cheaper than growing a region, as it doesn't require enumerating edges in the detector graph.

\subsubsection{An example}

\begin{figure}
    \centering
    \includegraphics[width=0.8\columnwidth]{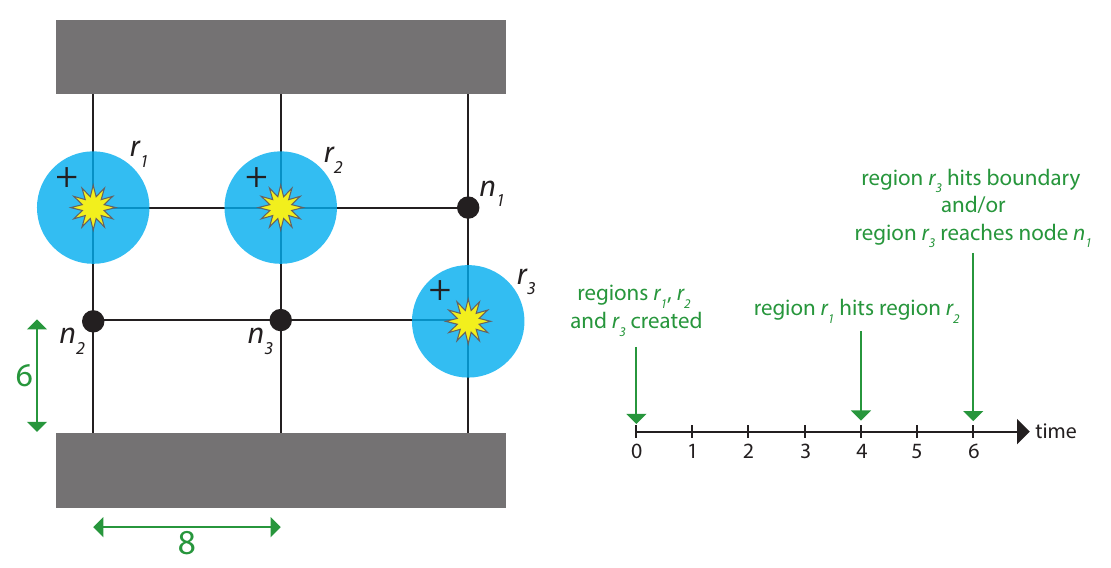}
    \caption{An example of some flooder events in a detector graph with three growing regions.}
    \label{fig:flooder_events}
\end{figure}

We give a small example of how the timeline of the flooder progresses in \fig{flooder_events}.
Since regions $r_1$, $r_2$ and $r_3$ are all initialised at $t=0$, their radius equations are all equal to $t$.
Regions $r_1$ and $r_2$ are separated by a single edge with weight $8$ and therefore collide at time $t=4$ (and recall that edge weights are always even integers to ensure collisions occur at integer times).
When the matcher is informed of the collision, $r_1$ and $r_2$ are matched and become frozen regions.
Region $r_3$ reaches empty node $n_1$ and the boundary at the same time ($t=6$), and so there are two equally valid sequences of events.
Either region $r_3$ matches to the boundary, and never reaches $n_1$, or $r_3$ reaches $n_1$ and then matches to the boundary.
Clearly the final state of the algorithm is not unique, however there is a unique solution \textit{weight}, and in this instance, both outcomes lead to the same set of compressed edges in the solution.

\subsubsection{Compressed tracking}

\begin{figure}
    \centering
    \includegraphics[width=0.9\columnwidth]{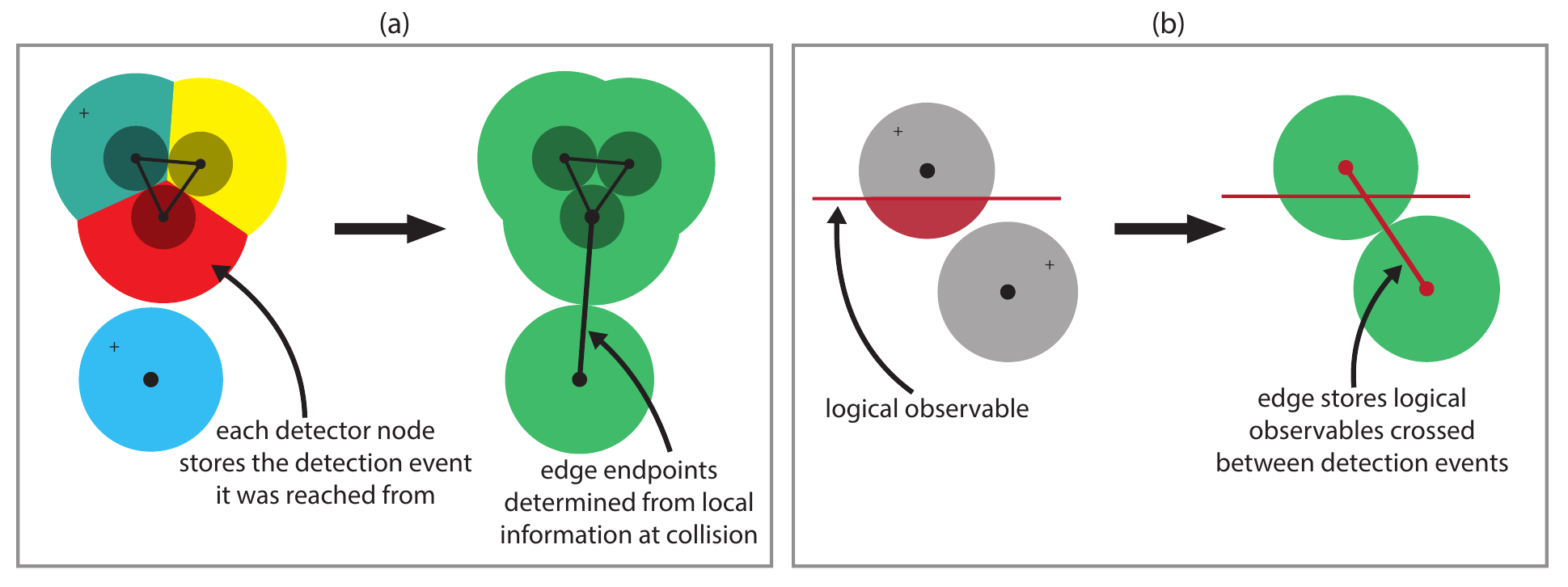}
    \caption{(a) As a region expands, each detector node it contains stores the detection event it was reached from (visualised by the Voronoi-style colouring of the blossom on the left). When a collision occurs, this allows the endpoints of the corresponding compressed edge (collision edge) to be determined from local information at the point of collision. (b) Each detector node also stores (as a 64-bit bitmask) the observables that were crossed to reach it from the detection event it was reached from. This allows the observables bitmask of the compressed edge to be recovered efficiently, also from local information at the point of collision.}
    \label{fig:compressed_tracking}
\end{figure}

Whenever a collision occurs between two regions $A$ and $B$, the flooder constructs a region edge $(A,B)$, which we recall includes a compressed edge corresponding to the shortest path between a detection event in $A$ and a detection event in $B$.
By storing relevant information on nodes as regions grow, the compressed edge can be constructed efficiently using local information at the point of collision (i.e.~using only information stored on the edge $(u, v)$ that the collision occurs on, and on the nodes $u$ and $v$).

This is explained in \fig{compressed_tracking}.
As a region $R$ reaches an empty node $v$ by growing along an edge $(u, v)$ from a predecessor node $u$, we store on $v$ a pointer to the detection event $\mathcal{S}(v)$ it was reached from (which can simply be copied from $u$).
In other words, we set $\mathcal{S}(v)\coloneqq \mathcal{S}(u)$ once $v$ is reached from $u$.
Initially, when a search is started from a detection event $w$ (i.e.~a trivial growing region is created containing $w$), then we set $\mathcal{S}(w)\coloneqq w$.
We refer to $\mathcal{S}(v)$ as the \textit{source detection event} of $v$.

We also store on $v$ the set of observables $l(v)$ crossed during the region growth (i.e.~the observables crossed along a shortest path from $\mathcal{S}(v)$ to $v$).
This set of crossed observables $l(v)$ can be efficiently computed when $v$ is reached from $u$ along edge $(u,v)$ from $l(v)\coloneqq l(u)\oplus l(u,v)$, where we implement $\oplus$ as a bitwise XOR since $l(u)$ and $l(u,v)$ are stored as bitmasks.
Initially, when a trivial growing region is created at a detection event $w$ we set $l(w)$ to the empty set.

Therefore, when a collision occurs between regions $R$ and $R^\prime$ along an edge $(p,q)$, the endpoints $(x,y)$ of the compressed edge associated with the collision edge $(R_x,R^\prime_y)$ can then be determined locally as $x=\mathcal{S}(p)$ and $y=\mathcal{S}(q)$.
The observables $l(x,y)$ associated with the collision edge can be computed locally as $l(x,y)\coloneqq l(p)\oplus l(q)\oplus l(p,q)$.
Note that compressed tracking can also be used to remove the peeling step of the union-find decoder~\cite{Delfosse2021almostlineartime}, as we explain in \app{compressed_uf}.
We note that it would also be interesting to explore how compressed tracking could be generalised to improve decoding for other families of codes that are not decodable with matching (for which the corresponding error models are not graphlike).

In PyMatching, we only fully rely on compressed tracking when there are 64 logical observables or fewer.
When there are more than 64 logical observables, we don't include the observable bitmask in the compressed edges, instead only storing the detection events at its endpoints. 
This still allows us to use sparse blossom to find which detection events are matched to each other.
Then after sparse blossom has completed, for each matched pair $(u, v)$ we use a bi-directional Dijkstra search (implemented by adapting the flooder and tracker as required) to find the shortest path between $u$ and $v$.
If $C\subseteq \mathcal{E}$ is the set of all edges along the shortest paths found this way by the Dijkstra search post-processing, then the solution output by PyMatching is $\bigoplus_{e_i\in C}l(e_i)$.
Note that since we are only post-processing with Dijkstra rather than constructing the full path graph, this only adds a small relative overhead (typically less than 50\%) to the runtime.

\subsection{Tracker}

The tracker is responsible for ensuring flooder events occur in the correct order.
A simple approach one \textit{could} take to implement the tracker would just be to place every flooder event in a priority queue.
However, many of the potential flooder events are chaff. 
For example, when a region $R$ is growing, a flooder event would be added to the queue for each of its neighboring edges.
We say an edge $(u, v)$ is a neighbor of a region $R$ if $u$ is in $R$ and $v$ is not (or vice versa).
Along each neighboring edge, there will be an event either corresponding to the region growing into an empty node, or colliding with another region or boundary.
However, if the region becomes frozen or shrinking, then all of these remaining events will be invalidated.

To reduce this chaff, rather than adding every flooder event to a priority queue, the tracker instead adds \textit{look-at-node} and \textit{look-at-region} events to a priority queue.
The flooder just finds the time of the \textit{next} event at a node or region, and asks the tracker for a reminder to look back at that time.
As a result, at each node, we only need to add the next event to the priority queue.
The remaining potential flooder events along neighboring edges will not be added if they have become invalidated.

When the flooder reschedules a node, it finds the time of the next ARRIVE or COLLIDE event along a neighboring edge, and asks the tracker for a reminder to look back at that time (a \textit{look-at-node} event).
The flooder finds the time of the next LEAVE or IMPLODE event for a shrinking region by checking the top of the shell area stack, and asks the tracker for a reminder to look back at the region at that time (a \textit{look-at-region} event).
To further reduce chaff, the tracker only adds a \textit{look-at-node} or \textit{look-at-region} event to the priority queue if it will occur at an earlier time than an event already in the queue for the same node or region.
Once the tracker reminds the flooder to look back at a node or region, the flooder checks if it is still a valid event by recomputing the next event for the node or region, processing it if so.

\subsection{Comparison between blossom and sparse blossom}

In \tab{concepts} we summarise how some of the concepts in the traditional blossom algorithm translate into concepts in sparse blossom.
If the traditional blossom algorithm is run on the path graph $\bar{\mathcal{G}}[\mathcal{D}]$ using a multiple tree approach (and with all dual variables initialised to zero at the start), then a valid state of blossom at a particular stage corresponds to a valid state of sparse blossom for the same problem at the appropriate point in the timeline.
The dual variables in blossom define the region radiuses in sparse blossom (and these radiuses can be used to construct the corresponding exploratory regions).
Likewise the edges in the alternating trees, blossom cycles and matches in traditional blossom can all be translated into compressed edges in the corresponding entities in sparse blossom.
We note, however, that when multiple alternating tree events happen at the same time in sparse blossom, \textit{any} ordering for the processing of these events is a valid choice.
So just because we can translate a valid state of one algorithm to that of the other, does not imply that two implementations of the algorithms (or the same algorithm) will have the same sequence of alternating tree manipuations (indeed it is unlikely that they will).
This correspondence between sparse blossom run on the detector graph and traditional blossom run on a path graph, and the correctness of blossom itself for finding a MWPM, is one way understanding why sparse blossom correctly finds a MWEM in the detector graph.

\begin{table}[p]
\centering
\begin{tabular}{|p{0.46\linewidth} | p{0.46\linewidth} |} 
 \hline
 Blossom concepts (multiple tree approach, applied to $\bar{\mathcal{G}}[\mathcal{D}]$) & Sparse blossom concepts  \\ [0.5ex] 
 \hline\hline
 Dual variable $y_u$ (for a node $u\in\mathcal{D}$) or $y_S$ (for a set of nodes $S\subseteq \mathcal{D}$ of odd cardinality at least three) & Radius $y_R$ of a graph fill region $R$, an exploratory region containing nodes and edges in the detector graph including an odd number of detection events. \\ 
 \hline
 A tight edge $(u, v)$ between two nodes $u$ and $v$ in $\mathcal{D}$ that is in the matching or belongs to an alternating tree or a blossom cycle. Since $(u, v)$ is an edge in the path graph, its weight is the length of a shortest path between the two detection events $u$ and $v$ in the detector graph $\mathcal{G}$ (found using a Dijkstra search when the path graph was constructed). It is \textit{tight} since the dual variables associated with $u$ and $v$ result in zero slack as defined in \eq{slack}. & A compressed edge $(u, v)$ associated with a region edge, belonging to a match, alternating tree or blossom cycle. The compressed edge $(u,v)$ represents a shortest path between two detection events $u$ and $v$ in the detector graph $\mathcal{G}$. Associated with $(u, v)$ is the set of logical observables $l(u,v)$ flipped by flipping edges in $\mathcal{G}$ along the corresponding path between $u$ and $v$. The two regions in the corresponding region edge are touching (the edge is tight).\\
 \hline
 An edge $(u, v)$ between two nodes $u$ and $v$ in $\mathcal{D}$ that is \textit{not} tight. Its weight is the length of the shortest path between $u$ and $v$ in the detector graph $\mathcal{G}$, found using a Dijkstra search when constructing the path graph. For typical QEC problems, the vast majority of edges in $\bar{\mathcal{G}}[\mathcal{D}]$ never become tight, but for standard blossom they still must be explicitly constructed using a Dijkstra search. & There is no analogous data structure for this edge in sparse blossom. The shortest path between $u$ and $v$ is not currently fully explored by graph fill regions. This means that either the path has not yet been discovered by sparse blossom (and may never be), or perhaps it had previously been discovered (belonging to a region edge) but at least one of the regions owning $u$ or $v$ since shrunk (e.g.~it was matched and then became a shrinking region in an alternating tree).\\
 \hline
 In the dual update stage, update each dual variable $y_u$ of an outer node with $y_u\coloneqq y_u + \delta$ and update each dual variable $y_u$ of an inner node with $y_u\coloneqq y_u - \delta$. The variable $\delta\in\mathbb{R}_{\geq 0}
$ is set to the maximum value such that the dual variables remain a feasible solution to the dual problem. & As time $\Delta t$ passes, each growing region $R$ explores the graph and its radius $y_R$ increases by $\Delta t$ and each shrinking region $R$ shrinks in the graph and its radius $y_R$ decreases by $\Delta t$. Eventually at $\Delta t=\delta$ a collision or implosion occurs (one of the matcher events in \fig{matcher_events}), which must be handled by the matcher.\\
 \hline
  An edge $(u,v)$ between a node $u$ in an alternating tree $T$ and a node $v$ in another alternating tree $T^\prime$ becomes tight after a dual update. The path between the root of $T$ and the root of $T^\prime$ is augmented and all nodes in the two trees become matches. & A growing region $R$ of an alternating tree $T$ collides with a growing region $R^\prime$ of another alternating tree $T^\prime$. The collision edge $(R, R^\prime)$ is constructed from local information at the point of collision. $R$ is matched to $R^\prime$ with the collision edge $(R, R^\prime)$ as a match edge. All other regions in the trees also become matched. All regions in the two trees become frozen.\\
 \hline
   An edge $(u, v)$ between two outer nodes $u$ and $v$ in the same tree $T$ becomes tight after a dual update. This forms an odd-length cycle in $T$ which becomes a \textit{blossom}, which itself becomes an outer node in $T$. & Two growing regions $R$ and $R^\prime$ from the same alternating tree $T$ collide. The discovered collision edge $(R, R^\prime)$ as well as the regions and tree-edges along the path between $R$ and $R^\prime$ in $T$ become a blossom cycle in a newly formed blossom. The blossom starts growing, and is now a growing region in $T$. \\
 \hline
\end{tabular}
\caption{The correspondence between concepts in the standard blossom algorithm and concepts in sparse blossom. For the standard blossom algorithm we assume a ``multiple tree with fixed $\delta$'' approach is used, and further we assume that the algorithm is applied to the path graph $\bar{\mathcal{G}}[\mathcal{D}]=(\mathcal{D}, \bar{\mathcal{E}})$.}
\label{tab:concepts}
\end{table}

\section{Data structures}\label{sec:data_structures}

In this section, we outline the data structures we use in sparse blossom.
Each detector node $u$ stores its neighbouring edges in the detector graph as an adjacency list.
For each neighbouring edge $(u, v)$, we store its weight $w((u,v))$ as a 32-bit integer, the observables $l(u,v)$ it flips as a 64-bit bitmask, as well as its neighbouring node $v$ as a pointer (or as nullptr for the boundary).
Edge weights are discretised as \textit{even} integers, which ensures that all events (including two growing regions colliding) occur at integer times.

Each \textit{occupied} detector node $v$ stores a pointer to the source detection event it was reached from $\mathcal{S}(v)$, a bitmask of the observables crosssed $l(v)$ along the path from its source detection event and a pointer to the graph fill region $C(v)$ it is contained in (the region that arrived at the node).
There are some properties that we cache on each occupied node whenever the blossom structure changes, in order to speed up the process of finding the next COLLIDE or ARRIVE event along an edge. 
This includes storing a pointer to the \textit{active} region the occupied detector node $v$ is owned by (the topmost blossom ancestor of the region it is contained in) as well as its radius of arrival $r_a^v$, which is the radius that the node's containing region $C(v)$ had when it arrived at $v$ (see \sect{reschedule_node}).
Additionally, we cache the \textit{wrapped radius} $r_w^v$ of each detector node $v$ whenever its owning regions' blossom structure changes (if a blossom is formed or shattered).
The wrapped radius of an occupied node $v$ is the local radius of the node, excluding the (potentially growing or shrinking) radius of the active region it is owned by.
If we let $y_A(t)$ be the radius of the active region that owns $v$, we can recover the local radius from the wrapped radius with $r_L^v(t)=r_w^v + y_A(t)$.

Each graph fill region $R$ has a pointer to its blossom parent and its topmost blossom ancestor. Its blossom cycle is stored as an array $L$ of \textit{cycle edges}, where the cycle edge $L[i]$ stores a pointer to the $i$th blossom child of $R$, along with the compressed edge associated with the region edge joining child $i$ to child $i+1\mod n_c$, where $n_c$ is the number of blossom children of $R$.
Its shell area stack is an array of pointers to the detector nodes it contains (in the order they were added).
For its radius $y_R(t)=mt+c$ we use 62 bits to store the $y$-intercept $c$ and with 2 bits used to store the gradient $c\in\{-1, 0, 1\}$.
Each region also stores its match as a pointer to the region it is matched to, along with the compressed edge associated with the match edge.
Finally, each growing or shrinking region has a pointer to an AltTreeNode.

An AltTreeNode is used to represent the structure of an alternating tree.
Each AltTreeNode corresponds to a growing region in the tree, as well as its shrinking parent region (if it has one).
Each AltTreeNode has a pointer to its growing graph fill region and its shrinking graph fill region (if it has one).
It also has a pointer to its parent AltTreeNode in the alternating tree (as a pointer and a compressed edge), as well as to its children (as an array of pointers and compressed edges).
We also store the compressed edge corresponding to the shortest path between the growing and shrinking region in the AltTreeNode.

Each detector node and graph fill region also has a tracker field, which stores the desired time the node or region should next be looked at, as well as the time (if any) it is already scheduled to be looked at as a result of a \textit{look-at-node} or \textit{look-at-region} event already in the priority queue (called the \textit{queued time}).
The tracker therefore only needs to add a new \textit{look-at-node} or \textit{look-at-region} event to the priority queue if its desired time is set to be \textit{earlier} than its current queued time.

A property of the algorithm is that each event dequeued from the priority queue must have a time that is greater than or equal to all previous events dequeued.
This allows us to use a radix heap~\cite{ahuja1990faster}, which is an efficient monotone priority queue with good caching performance.
Since the next flooder event can never be more than one edge length away from the current time, we use a cyclic time window for the priority, rather than the total time.
We use 24, 32 and 64 bits of integer precision for the edge weights, flooder event priorities and total time, respectively.

\section{Expected running time}\label{sec:expected_running_time}

\begin{figure}
\centering
     \begin{subfigure}[b]{0.48\textwidth}
         \centering
         \includegraphics[width=\textwidth]{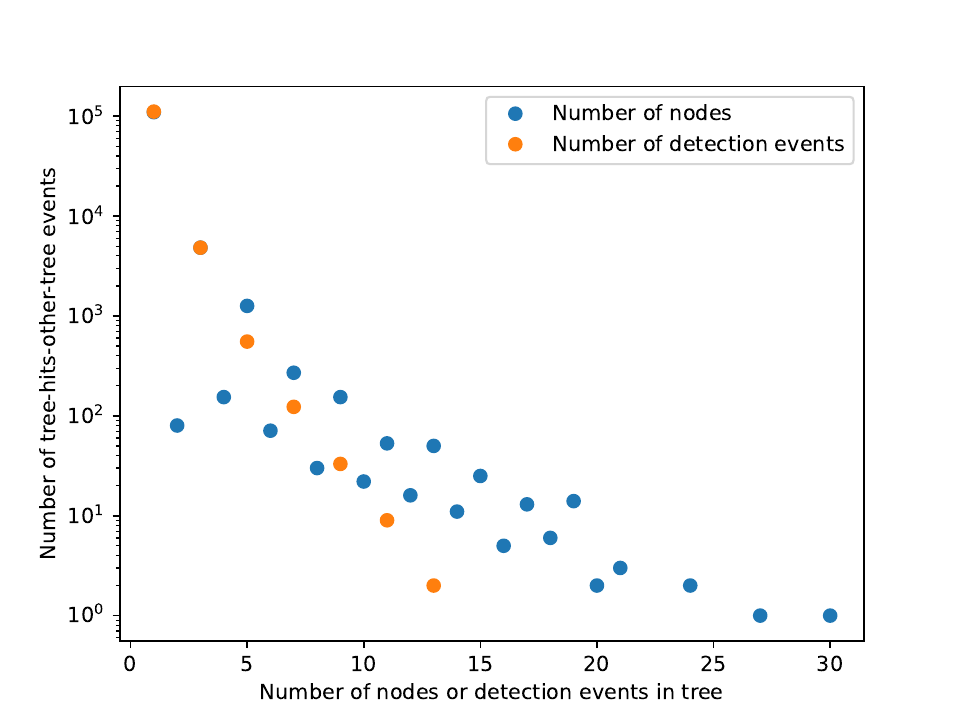}
         \caption{}
         \label{fig:alternating_tree_size}
     \end{subfigure}
     \hfill
     \begin{subfigure}[b]{0.48\textwidth}
         \centering
         \includegraphics[width=\textwidth]{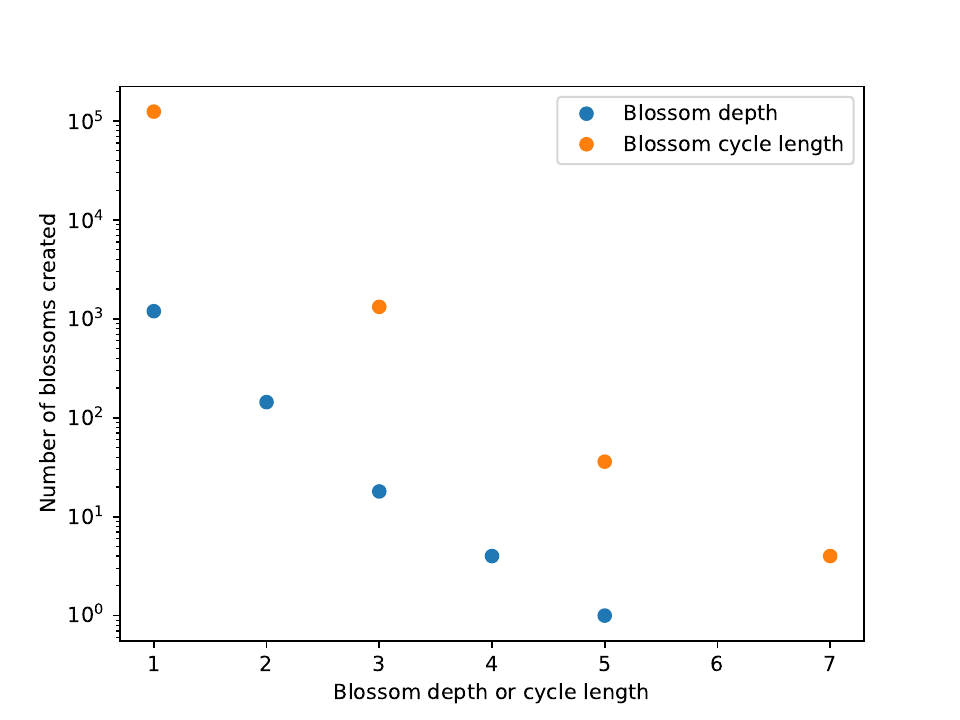}
         \caption{}
         \label{fig:blossom_size}
     \end{subfigure}
     \caption{Distribution of alternating tree and blossom sizes observed in sparse blossom when decoding 1000 shots of distance-11 surface code circuits with $p=0.3\%$ circuit-level noise. (a) A histogram of the size of alternating trees observed in events where a tree hits another tree, in terms of both the number of detection events and detector nodes contained in each tree. (b) A histogram of the size of blossoms formed, in terms of both the length of each blossom's blossom cycle, as well as its recursive depth. A blossom depth or cycle of size one is a trivial blossom (a graph fill region without blossom children).}
     \label{fig:exponential_decay}
\end{figure}

Empirically, we observe an almost-linear running time of our algorithm for surface codes below threshold (see \fig{timing_p1000}).
We would expect the running time to be linear at low physical error rates, since in this regime a typical error configuration will consist of small isolated clusters of errors.
Provided the clusters are sufficiently well separated from one another, each cluster is essentially handled as an independent matching problem by sparse blossom.
Furthermore, using results from percolation theory~\cite{menshikov1986coincidence}, we would expect the number of clusters of a given size to decay exponentially in this size~\cite{fowler2013minimum}.
Since the number of operations required to match a cluster is polynomial in its size, this leads to a constant cost per cluster, and therefore an expected running time that is linear in the size of the graph.

In more detail, suppose we highlight every edge in the detector graph $\mathcal{G}$ ever touched by sparse blossom when decoding a particular syndrome.
Now consider the subgraph $\mathcal{F}$ of $\mathcal{G}$ induced by these highlighted edges.
This graph will generally have many connected components, and we will refer to each connected component as a \textit{cluster region}.
Clearly, running sparse blossom on each cluster region separately will give an identical solution to running the algorithm on $\mathcal{G}$ as a whole.
Let us \textit{assume} that the probability that a detector node in $\mathcal{G}$ is within a cluster region of $n_c$ detector nodes is at most $Ae^{-bn_c}$ for some $b>0$.
Since the worst-case running time of sparse blossom is polynomial in the number of nodes $n$ (at most $O(n^4)$, see \app{worst_case}), the expected running time to decode a cluster region (if any) at a given node is at most $\sum_{n_c=1}^\infty Ae^{-bn_c}n_c^4=O(1)$, i.e.~constant.
Therefore, the running time is linear in the number of nodes.
Here we have assumed that the probability of observing a cluster region at a node decays exponentially in its size.
However, in \cite{fowler2013minimum} it was shown that this is indeed the case at very low error rates.
Furthermore, we provide empirical evidence for this exponential decay for error rates of practical interest in \fig{exponential_decay}, where we plot the distribution of the sizes of alternating trees and blossoms observed when using sparse blossom to decode surface codes with 0.3\% circuit-level noise.
Our benchmarks in \fig{timing_p1000} and \fig{timing_p200} are also consistent with a running time that is roughly linear in the number of nodes, and even above threshold (\fig{timing_p100}) the observed complexity of $O(n^{1.32})$ is only slightly worse than linear.

Note that here we have ignored the fact that our radix heap priority queue is shared by all clusters.
This does not impact the overall theoretical complexity, since the insert and extract-min operations of the radix heap have an amortized time complexity that is independent of the number of items in the queue.
In particular, inserting an element takes $O(1)$ time, and extract-min takes amortized $O(B)$ time, where $B$ is the number of bits used to store the priority (for us $B=32$).

Nevertheless, our use of the timeline concept and a multiple tree fixed $\delta$ approach (relying on the priority queue) does result in more cache misses for very large problem instances, empirically resulting in an increase in the processing time per detection event.
This is because events that are local in time (and therefore processed soon after one another) are in general not local in memory, since they can require inspecting regions or edges that are very far from one another in the detector graph.
In contrast we would expect a single tree approach to have better memory locality for very large problem instances.
However, an advantage of the multiple tree approach we have taken is that it ``touches'' less of the detector graph.
For example, consider a simple problem of two isolated detection events in a uniform detector graph, separated by distance $d$.
In sparse blossom, since we use a multiple tree approach, these two regions will grow at the same rate and will have explored a region of radius $d/2$ when they collide.
 In a 3D graph, let's assume that a region of radius $r$ touches $\approx kr^3$ edges for some constant $k$.
 In contrast, in a single tree approach (such as used by Refs.~\cite{fowler2012towards, fowler2012timing_analysis, fowler2013minimum}), one region will grow to radius $d$ and collide with the region of the other detection event, which will still have radius 0.
 Therefore, the multiple tree approach has touched $\approx 2k(d/2)^3$ edges, $\approx 4\times$ fewer than the $kd^3$ edges touched by the single tree approach.
This is essentially the same advantage that a bidirectional Dijkstra search has over a regular Dijkstra search.

\section{Computational results}\label{sec:computational_results}

\begin{figure}
    \centering
    \includegraphics[width=0.7\columnwidth]{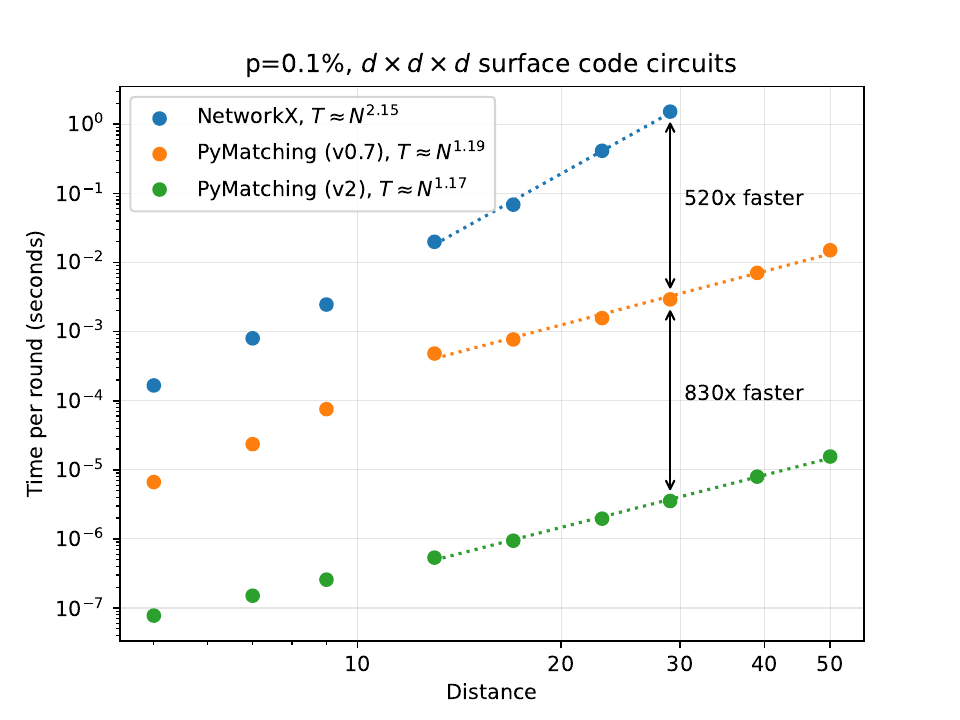}
    \caption{Decoding time per round for PyMatching v2 (our implementation of sparse blossom), compared to PyMatching v0.7 and a NetworkX implementation.
    For distance $d$, we find the time to decode $d$ rounds of a distance $d$ surface code circuit and divide this time by $d$ to obtain the time per round.
    We use a circuit-level depolarising noise model where the probability $p=0.1\%$ sets the strength of two-qubit depolarising noise after each CNOT gate, the probability that each reset or measurement fails, as well as the strength of single-qubit depolarising noise applied before each round.
    The threshold for this noise model is around $0.71\%$.
    PyMatching v0.7 uses a C++ implementation of the local matching algorithm described in~\cite{higgott2022pymatching}. The pure Python NetworkX implementation first constructs a complete graph on the detection events, where each edge $(u, v)$ represents a shortest path between $u$ and $v$, and then uses the standard blossom algorithm on this graph to decode.
    All three decoders use a single core of an M1 Max processor.}
    \label{fig:timing_p1000}
\end{figure}

\begin{figure}
    \centering
    \includegraphics[width=0.7\columnwidth]{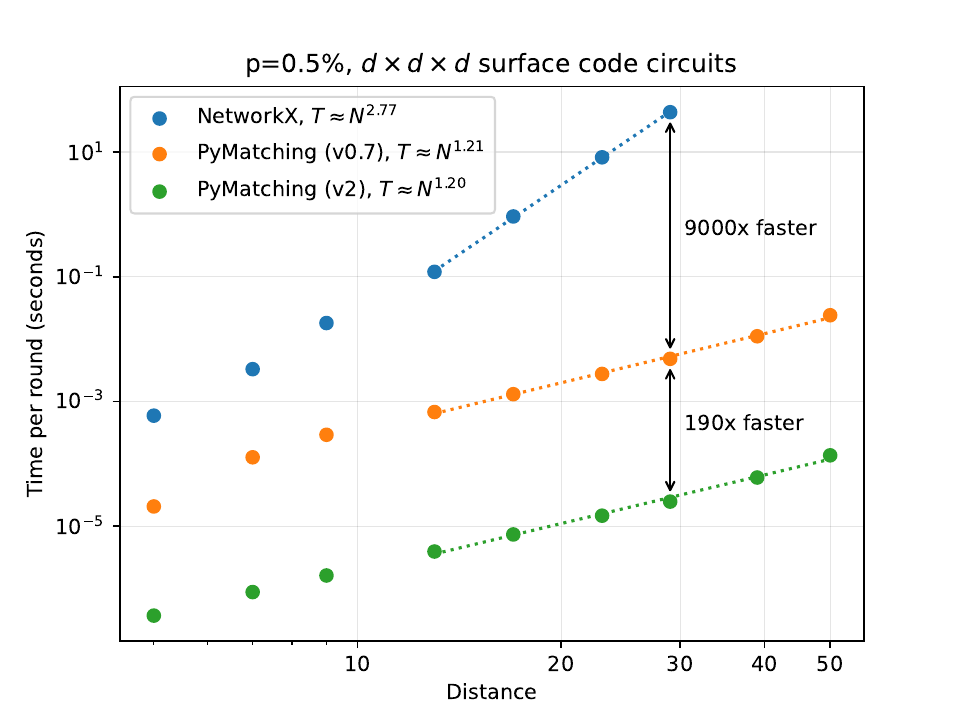}
    \caption{Decoding time per round for PyMatching v2 (our implementation of sparse blossom), compared to PyMatching v0.7 and a NetworkX implementation.
    The only difference with \fig{timing_p1000} is that here we set $p=0.5\%$.}
    \label{fig:timing_p200}
\end{figure}

\begin{figure}
    \centering
    \includegraphics[width=0.7\columnwidth]{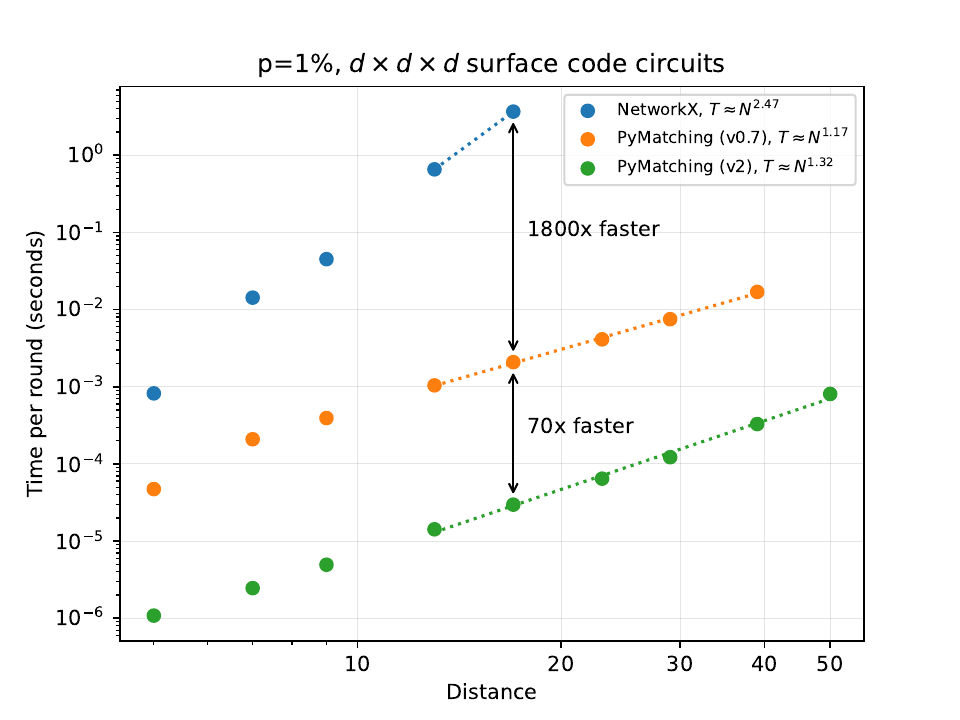}
    \caption{Decoding time per round for PyMatching v2 (our implementation of sparse blossom), compared to PyMatching v0.7 and a NetworkX implementation.
    The only difference with \fig{timing_p1000} is that here we set $p=1\%$.}
    \label{fig:timing_p100}
\end{figure}

We benchmarked the running time of our implementation of sparse blossom (PyMatching 2) for decoding surface code memory experiments (see \fig{timing_p1000}, \fig{timing_p200} and \fig{timing_p100}).
For 0.1\% circuit-level depolarising noise, sparse blossom processes both X and Z bases of distance-17 surface code circuits in less than one microsecond per round of syndrome extraction on a single core, which matches the rate at which syndrome data is generated by superconducting quantum computers.

At low physical error rates (e.g.~\fig{timing_p1000}), the roughly linear scaling of PyMatching v2 is a quadratic improvement over the empirical scaling of an implementation that constructs the path graph explicitly and solves the traditional MWPM problem as a separate subroutine.
For 0.1\%-1\% physical error rates and distance 29 and larger, PyMatching v2 is $>100,000\times$ faster than a pure Python implementation that uses the exact reduction to MWPM.
Compared to the local matching approximation of the MWPM decoder used in \cite{higgott2022pymatching}, PyMatching v2 has a similar empirical scaling but is around $100\times$ faster near threshold (error rates of 0.5\% to 1\%) and almost $1000\times$ faster below threshold ($p=0.1\%$).
Much of this speedup relative to local matching is due to the fact that local matching still explicitly constructs much more of the path graph than is ultimately used by the blossom algorithm.
This is because the truncation of the path graph in local matching is not done adaptively during the blossom algorithm but instead in a separate step before the blossom algorithm begins.
Furthermore, local matching is an approximation of the MWPM decoder, in contrast to Sparse Blossom (which is exact).

\begin{figure}
\centering
     \begin{subfigure}[b]{0.48\textwidth}
         \centering
         \includegraphics[width=\textwidth]{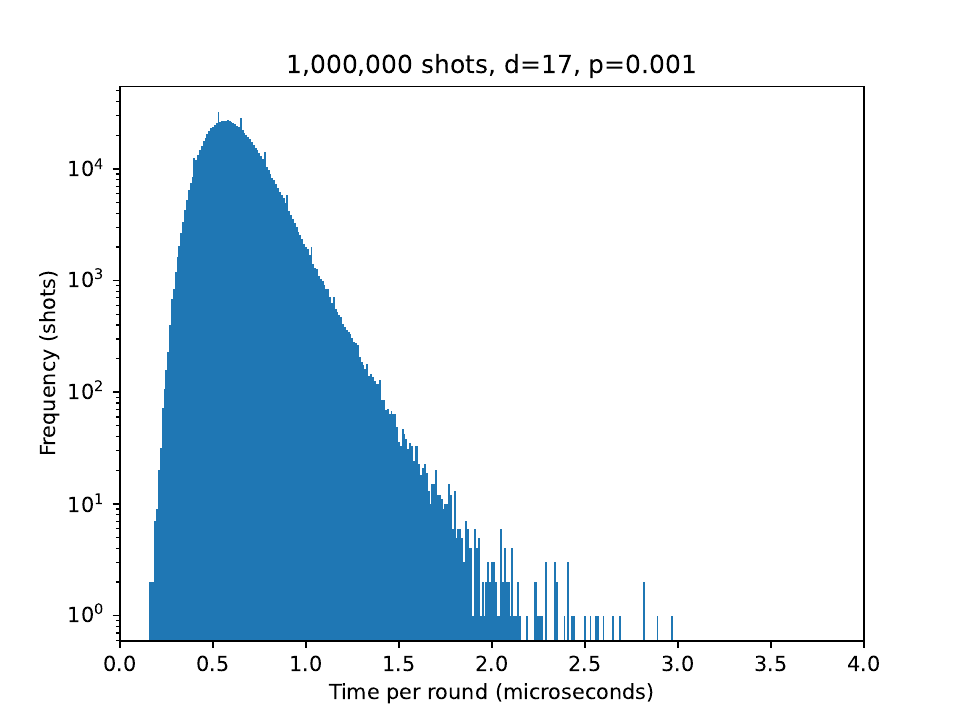}
         \caption{}
         \label{fig:time_per_round_distribution_d17_p1000}
     \end{subfigure}
     \begin{subfigure}[b]{0.48\textwidth}
         \centering
         \includegraphics[width=\textwidth]{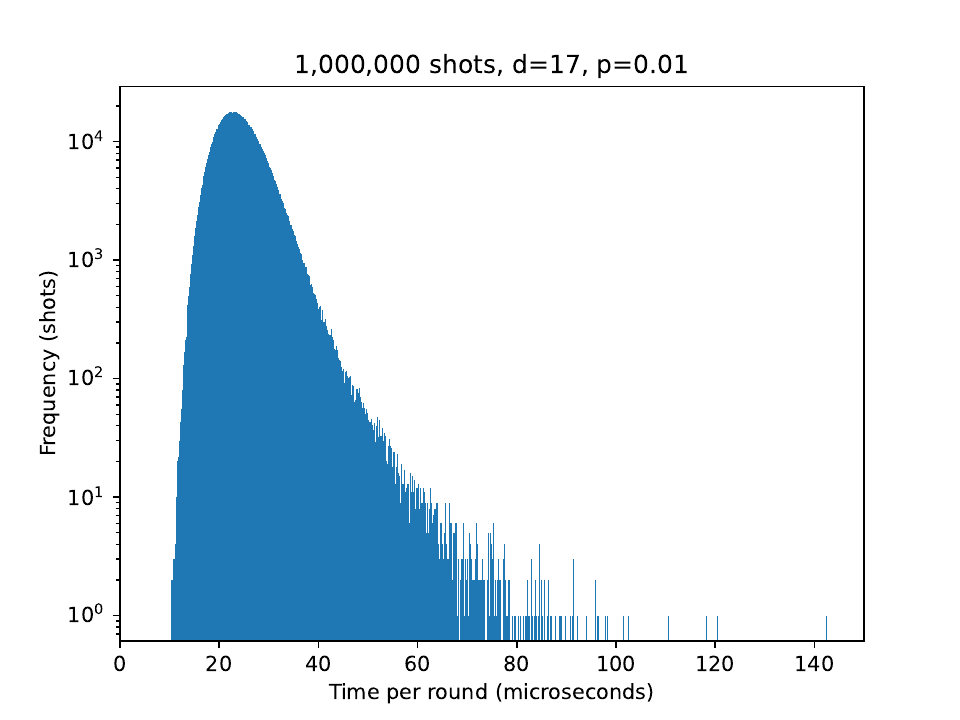}
         \caption{}
         \label{fig:time_per_round_distribution_d17_p100}
     \end{subfigure}
     \caption[Distribution of sparse blossom running times]{Histograms showing the distribution of running times of sparse blossom using 17-round, distance-17 surface code circuits and a standard circuit-level depolarising noise model. In (a) we use $p=0.1\%$ and a histogram bin width of 0.01 microseconds. 97.4\% of the shots have a running time per round below 1 microsecond. In (b) we instead use $p=1\%$ and a histogram bin width of 0.2 microseconds.}
     \label{fig:distribution_of_runtimes}
\end{figure}

\begin{figure}
    \centering
    \includegraphics[width=0.6\textwidth]{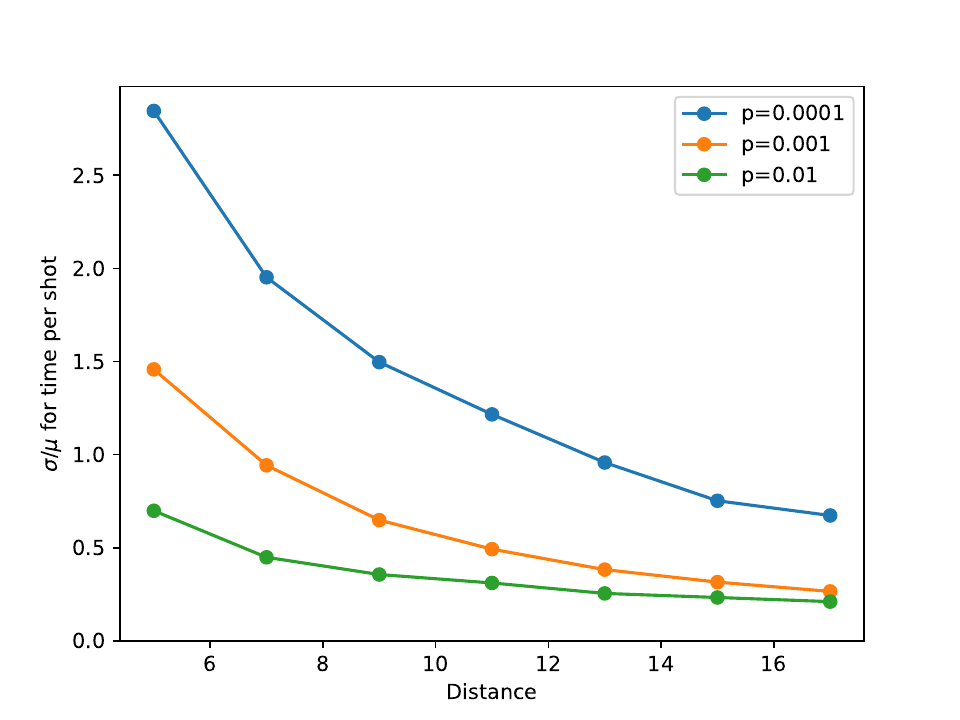}
    \caption[Relative standard deviation $\sigma/\mu$ of running time]{Relative standard deviation $\sigma/\mu$ of the time per shot for distance-$d$ surface code circuits (with $d$ rounds) and standard circuit-level depolarising noise. Here $\sigma$ and $\mu$ are the standard deviation and mean, respectively, of the time per shot, sampling 1 million shots for each data point.}
    \label{fig:relative_standard_deviation_time_per_shot}
\end{figure}

We analysed the distribution of the running time per shot of PyMatching v2 for simulated surface code data, see \fig{distribution_of_runtimes} and \fig{relative_standard_deviation_time_per_shot}.
For example, for distance-17 surface code circuits with $p=0.1\%$ circuit-level noise, we observe a mean running time of $0.62$ microseconds per round and find that 97.4\% of the million shots were decoded with a running time below 1 microsecond per round.
We also plot the relative standard deviation $\sigma/\mu$ of the running time per shot in \fig{relative_standard_deviation_time_per_shot} and find that $\sigma/\mu$ decreases as either the distance or error rate is increased.

We also compared the speed of PyMatching v0.7 with that of PyMatching v2 on experimental data, by running both decoders on the full dataset of Google's recent experiment demonstrating the suppression of quantum errors by scaling a surface code logical qubit from distance 3 to distance 5~\cite{google2023suppressing,google_quantum_ai_team_2022_6804040}.
On an M2 chip, PyMatching v0.7 took 3 hours and 43 minutes to decode all 7 million shots in the dataset, whereas PyMatching v2 took 71 seconds.

\section{Conclusion}\label{sec:conclusion}

In this work, we have introduced a variant of the blossom algorithm, which we call sparse blossom, that directly solves the minimum-weight perfect matching decoding problem relevant to error correction.
Our approach avoids the computationally expensive all-to-all Dijkstra searches often used in implementations of the MWPM decoder, where a reduction to the traditional blossom algorithm is used.
Our implementation, available in version 2 of the open-source PyMatching Python package, can process around a million errors per second on a single core.
For a distance-17 surface code, it can decode both $X$ and $Z$ bases in under $1\si\us$ per round of error correction, which matches the rate at which syndrome data is generated on a superconducting quantum computer.

Some of the techniques we have introduced can be directly applied to improve the performance of other decoders.
For example, we introduced \textit{compressed tracking}, which exploits the fact that the decoder only needs to predict which \textit{logical observables} were flipped, rather than the physical errors themselves.
This allowed us to use a sparse representation of paths in the detector graph, storing only the endpoints of a path, along with the logical observables it flips (as a bitmask).
We showed that compressed tracking can be used to significantly simplify the union-find decoder (see \app{compressed_uf}), leading to a compressed representation of the disjoint-set data structure and eliminating the need to construct a spanning tree in the peeling step of the algorithm.

When used for error correction simulations, our implementation can be trivially parallelised across batches of shots.
However, achieving the throughput necessary for real-time decoding at scale motivates the development of a parallelised implementation of sparse blossom.
For example, for the practically relevant task of decoding a distance-30 surface code at 0.1\% circuit-level noise, the throughput of sparse blossom is around $3.5\times$ slower than the $<1\si\us$ per round throughput desired for a superconducting quantum computer.
It would therefore be interesting to investigate whether a multi-core CPU or FPGA-based implementation could achieve the throughput necessary for real-time decoding at scale by adapting techniques in \cite{wu2023fusion} for sparse blossom.
Finally, it would be interesting to adapt sparse blossom to exploit correlated errors that arise in realistic noise models, for example by using sparse blossom as a subroutine in an optimised implementation of a correlated matching~\cite{fowler2013optimal} or belief-matching~\cite{higgott2022fragile} decoder.

\section*{Contributions}

Both authors worked together on the design of the sparse blossom algorithm. Oscar Higgott wrote the paper and the majority of the code. Craig Gidney guided the project and implemented some final optimisations such as the radix heap.

\section*{Acknowledgements}

We would like to thank Nikolas Breuckmann, Dan Browne, Austin Fowler, Noah Shutty and Barbara Terhal for helpful feedback that improved the manuscript.

\printbibliography

\appendix

\section{Compressed tracking in Union-Find}\label{app:compressed_uf}

\begin{figure}
    \centering
    \includegraphics[width=0.8\columnwidth]{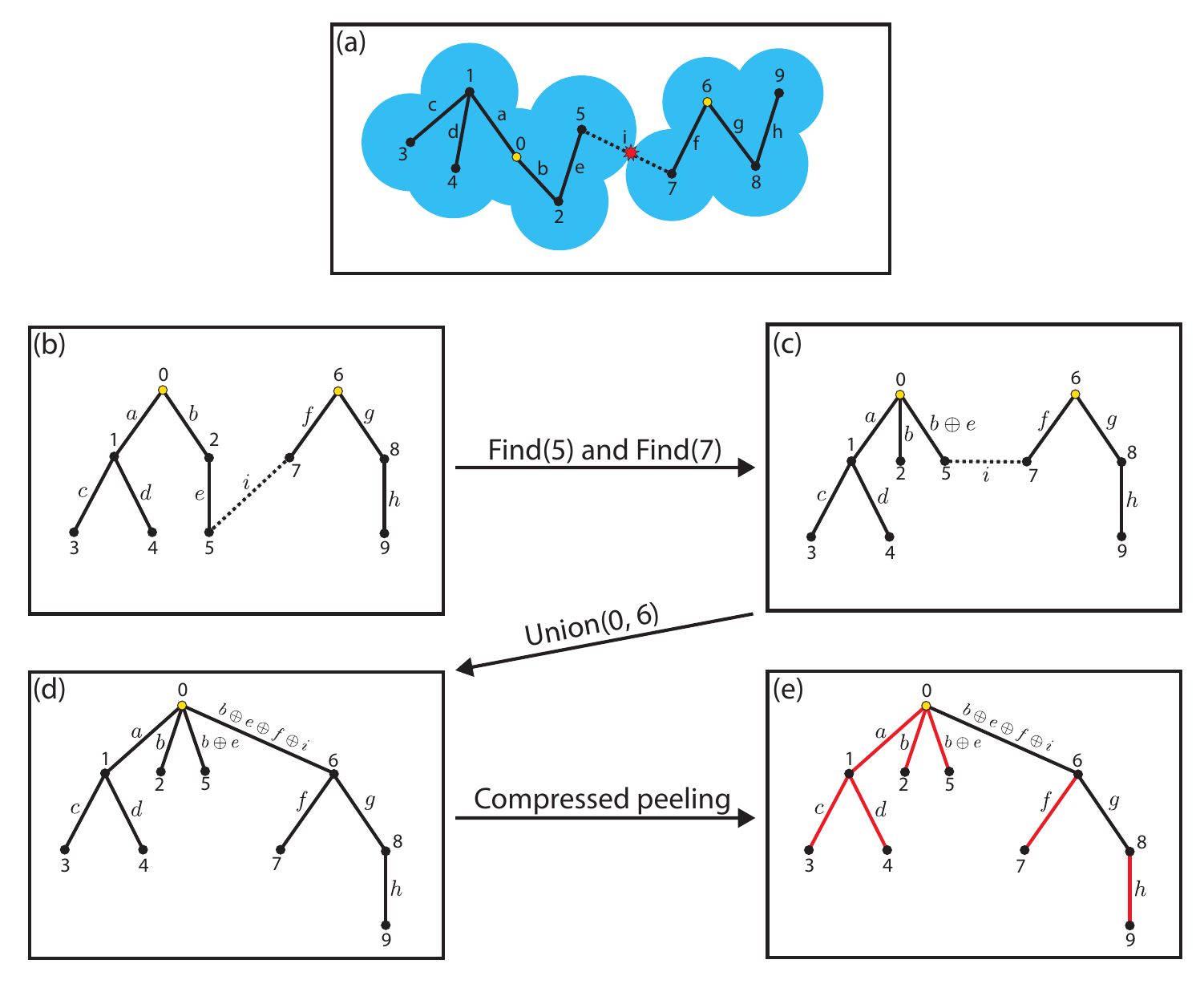}
    \caption{Compressed tracking in Union-Find. (a) Two clusters collide. Each node in the cluster tree denotes a detection event, and each edge is a compressed edge representing a path between the two detection events in the detector graph (not necessarily a shortest path). Each edge is labelled with a letter denoting the bitmask of the observables crossed along the path it represents. This differs from traditional Union-Find implementations, where every detector node in a cluster is a node in the cluster tree. When two clusters collide, we store a compressed edge for the path between detection events along which the collision occurred (the collision edge). (b) The two cluster trees, along with the collision edge (dashed line). (c) After finding the source detection events (5 and 7) involved in the collision, we call Find(5) and Find(7). When using path compression (which here connects node 5 to the root node), we ensure the observable bitmask is kept up-to-date by taking the sum (modulo 2) of bitmasks along the path to the root node. (d) Union(0, 6) adds the smaller cluster as a subtree of the root node of the larger cluster (node 6 becomes a child of node 0). We store the observable bitmask along edge (0, 6), by taking the sum (modulo 2) of the observable bitmasks on edges (0, 5), (6, 7) and the collision edge (5, 7). (e) The combined cluster now has even parity. We use compressed peeling to highlight a set of edges (shown in red) such that each node is incident to an odd number of edges. We take the sum (modulo 2) of the observable bitmasks on these highlighted edges to find the predicted logical observable.}
    \label{fig:union_find}
\end{figure}

Compressed tracking can be naturally adapted to the Union-Find decoder~\cite{Delfosse2021almostlineartime, huang2020fault}, as shown in \fig{union_find}.
Each detection event is initialised with a region, which we refer to as a cluster, to be consistent with Ref.~\cite{Delfosse2021almostlineartime}.
Using the same approach as for the compressed tracking in sparse blossom, we can track the detection event that each detector node has been reached from, as well as the observables that have been crossed to reach it.
More explicitly, let $\mathcal{S}(u)$ again denote the \textit{source detection event} of a detector node $u$, and we initialise $\mathcal{S}(x)=x$ for each detection event $x$ at the start of the algorithm.
We denote the set of logical observables crossed an odd number of times from a node's source detection event by $l(u)$, and we initialise $l(x)$ as the empty set for each detection event $x$.
We grow a cluster $C$ by adding a node $v$ from an edge $e\coloneqq(u, v)$ on the boundary of $C$ (i.e.~an edge such that $u\in C$ and $v\notin C$).
As we add $v$ to $C$, we set $\mathcal{S}(v)=\mathcal{S}(u)$ and $l(v)=l(u)\oplus l(e)$.
Recall that $l(e)$ is the set of logical observables flipped by edge $e$ and $\oplus$ denotes the symmetric difference of sets.
Since we store $l(e_i)$ as a bitmask for each edge $e_i\in\mathcal{E}$, the symmetric difference of two edges $l(e_i)\oplus l(e_j)$ can be implemented particularly efficiently using a bitwise XOR.

We represent each cluster as a \textit{compressed cluster tree}.
Each node in the compressed cluster tree corresponds to a detection event, in contrast to the cluster tree introduced in \cite{Delfosse2021almostlineartime}, where each node is a detector node.
Each edge $c\coloneqq(x, y)$ in the cluster tree is a compressed edge, representing a path $P$ through the detector graph between detection events $x$ and $y$.
In contrast to compressed edges in sparse blossom, this path is in general not the minimum-length path.
The compressed edge $c$ is assigned the logical observables crossed an odd number of times by $P$, denoted $l(c)$ or $l(x,y)$.
We check which cluster a detector node $u$ belongs to by calling $\mathrm{Find}(\mathcal{S}(u))$.

We modify the path compression step of the Find operation such that whenever a path $B$ (consisting of compressed edges) through the cluster tree between two detection events $f$ and $g$ is replaced by a compressed edge $c\coloneqq (f, g)$, the set of logical observables $l(c)$ of the new compressed edge is calculated $l(c)\coloneqq \bigoplus_{c_i\in B}l(c_i)$.
A similar modification can be made if path splitting or path halving is used instead of path compression for the Find operation.

The Union operation is adapted in a similar manner.
Suppose a cluster $C_i$ collides with a cluster $C_j$ along an edge $(u, v)$.
In other words, $(u, v)$ is an edge in the detector graph such that $u\in C_i$ and $v\in C_j$.
We construct the collision edge $(\mathcal{S}(u), \mathcal{S}(v))$ from local information at the point of collision, and assign it the set of logical observables $l(\mathcal{S}(u), \mathcal{S}(v))\coloneqq l(u)\oplus l(v) \oplus l(u, v)$.
When we merge cluster $C_i$ with cluster $C_j$ using the Union operation, we add the root node $\mathcal{R}(C_i)$ of the smaller cluster tree (say $C_i$) as a child of the root node $\mathcal{R}(C_j)$ of the larger cluster tree $C_j$ by adding a compressed edge $c_{ij}\coloneqq (\mathcal{R}(C_i),\mathcal{R}(C_j))$ to the tree.
We assign its set of logical observables $l(c_{ij})$ to be $l(c_{ij})\coloneqq \bigoplus_{c_k\in P_{ij}}l(c_k)$, where $P_{ij}$ is the path through the tree between $\mathcal{R}(C_i)$ and $\mathcal{R}(C_j)$.

Finally, once all clusters have even parity (an even number of detection events, or connected to the boundary), we can apply the peeling algorithm~\cite{delfosse2020linear} to the compressed cluster trees, which returns a set of compressed edges $\mathcal{P}$. 
We say the compressed edges in $\mathcal{P}$ are \textit{highlighted} edges in the cluster tree.
The set of logical observables that the decoder predicts to have been flippsed is then $\bigoplus_{c_i\in\mathcal{P}}l(c_i)$.
This stage is much more efficient than the traditional peeling step of UF, as we do not need to construct a spanning tree in each cluster.
Instead, we only run peeling on our compressed representation of the cluster trees.
Compressed peeling is linear in the number of compressed edges in the cluster tree.
For completeness, we give pseudocode for compressed peeling in \Cref{alg:compressed_peeling}, however this is simply a recursive definition of the peeling algorithm of \cite{delfosse2020linear} applied to the case of a compressed cluster tree comprised of a graph of compressed edges joining detection events, rather than to a spanning tree subgraph of a conventional union-find cluster in the detector graph (comprised of detector nodes and detector graph edges).
The procedure is recursive and takes as input a node $x$ in the compressed cluster tree, returning $p_x$ and $l_x$.
Here, $l_x$ is the set of logical observables flipped by flipping the highlighted compressed edges that are descendants of $x$ in the cluster tree (a descendant edge of $x$ is an edge in the subtree rooted at $x$).
The auxiliary variable $p_x$ (used in the recursion) is the parity of the number of highlighted child edges of $x$ in the cluster (where a child edge is an edge connecting $x$ to one of its children in the cluster tree).
Initially, we assume no node in the cluster tree is connected to the boundary, in which case we initialise $p_x^{init}=even$ for each node $x$.
We run compressed peeling on the root node $r$ to find the set of logical observables $l_r$ flipped by all highlighted edges in the cluster tree.
Note that $p_r$ should always be odd for the root node if there is no boundary.

Recall that peeling should find a set of highlighted edges (edges in $\mathcal{P}$) such that each node in the tree is incident to an odd number of highlighted edges.
We will first show that $l_x$ (returned by compressed peeling) is indeed the set of logical observables flipped by highlighted edges that are descendants of $x$, and that $p_x$ is the parity of the number of highlighted child edges of $x$.
Consider the base case that $x$ is a leaf node, in which case it has no child edges or descendants.
In this case, compressed peeling correctly sets $p_x$ to \textit{even} (since we initialise $p_x^{init}=even$) and $l_x$ to the empty set.
Now consider the inductive step, where $x$ has children $\mathcal{C}(x)$ in the cluster tree.
For each $y\in \mathcal{C}(x)$ we highlight the edge $(x,y)$ if $p_y$ is even, and $p_x$ is set to the parity of the number of these highlighted child edges of $x$, as required.
The main loop of the algorithm sets $l_x$ to $$l_x=\left(\bigoplus_{y\in\mathcal{C}(x)}l_y\right)\oplus \left(\bigoplus_{(x,w): w \in \mathcal{C}(x), (x,w)\in \mathcal{P}}l(x,w)\right)$$
and if we assume $l_y$ is the set of logical observables flipped by highlighted descendant edges of $y$ then clearly $l_x$ is the set of logical observables flipped by highlighted descendant edges of $x$.
Finally, note that since we apply compressed peeling to a tree, the function is called on each node $x$ exactly once, and we highlight an edge $(x,y)$ to a child $y$ of $x$ if and only if $p_y$ is even.
Therefore, each node becomes incident to an odd number of highlighted edges, as required.

We haven't yet considered the boundary. If there is a compressed edge $(x, b)$ in a cluster tree $C$ connecting a detection event $x$ to the boundary $b$ (there can be at most one such edge since a cluster becomes neutral once it hits the boundary), we first add $l(x,b)$ to the solution and remove $(x,b)$ from $C$, then we set $p_x^{init}=odd$ before applying compressed peeling to the root node $r$ of the remaining cluster tree, adding the resulting $l_r$ to the solution.

\begin{algorithm}
\caption{Compressed Peeling}\label{alg:compressed_peeling}
\begin{algorithmic}
\Procedure{CompressedPeeling}{$x$}
  \State $p_x \gets p_x^{init}$
  \algorithmiccomment{Parity of the number of highlighted child edges of $x$}
  \State $l_x \gets \{\}$
  \algorithmiccomment{Observables flipped by highlighted descendant edges of $x$}
  \For{each child $y$ of $x$} 
  \State $p_y$, $l_y \gets$ \Call{CompressedPeeling}{$y$}
  \State $l_x \gets l_x \oplus l_y$
  \If {$p_y$ is even}
    \State $l_x \gets l_x \oplus l(x, y)$ \algorithmiccomment{Compressed edge $(x, y)$ becomes highlighted}
    \State \textit{flip} $p_x$
  \EndIf
  \EndFor
  \State \Return $p_x$, $l_x$
\EndProcedure
\end{algorithmic}
\end{algorithm}

\section{Worst-case running time}\label{app:worst_case}

In this section, we will find a worst-case upper bound of the running time of sparse blossom.
Note that this upper bound is likely loose, and furthermore differs greatly from the expected running time of sparse blossom for typical QEC problems, which we believe to be linear in the size of the graph.
Let us denote the number of detector nodes by $n$, the number of edges in the detector graph by $m$ and the number of detection events by $q$.
First, note that each alternating tree always contains exactly one unmatched detection event, in the sense that only a single growing region needs to become matched for the whole tree to shatter into matched regions.
Therefore, the alternating tree events that grow the alternating tree, form blossoms or shatter blossoms (events of type a, c, d or e in \fig{matcher_events}) do not change the number of detection events that remain to be matched.
On the other hand, when a tree hits another tree (type b), the number of unmatched detection events reduces by two, and when a tree hits the boundary (type f), or a boundary match (type g), then the number of unmatched detection events reduces by one.
We refer to an alternating tree event that reduces the number of unmatched detection events as an \textit{augmentation}, and refer to a portion of the algorithm between consecutive augmentations as a \textit{stage}.
Clearly, there can be at most $q$ augmentations and at most $q$ stages.

We now bound the complexity of each augmentation, and of each stage.
In each stage, there are at most $O(q)$ blossoms formed or shattered, and at most $O(q)$ matches added to trees, since the same blossom cannot form and then shatter within a stage.
To explain more concretely, first note that there are at most $O(q)$ blossoms or trivial regions at any moment in the algorithm.
Within a stage, the only alternating tree events that are allowed are of type a, c, d and e (since b, f and g events are augmentations).
Once a blossom is growing, the same blossom cannot become a shrinking region and shatter until after an augmentation, since the blossom must first become a match (through an augmentation) and then become an inner node in a type-a event.
Therefore at most $O(q)$ blossoms can form in each stage.
Finally, within each stage the only blossoms that can shatter, or be added to a tree as part of a match, are those blossoms (or trivial regions) that were already formed at the beginning of the stage, and there at most $O(q)$ of these.

We now consider the worst-case complexity of each of these possible operations within a stage.
When a blossom forms, each node it owns updates its cached topmost blossom ancestor and wrapped radius, with cost proportional to the depth of the blossom tree, and this step has complexity $O(nq)$.
Additionally, every node owned by the blossom is rescheduled, with $O(m)$ cost.
Updating the blossom structure and alternating tree structure (e.g.~the compressed edges) is at most $O(q)$.
In total, forming a blossom has a running time of at most $O(nq + m)$, and the same upper bound applies for shattering a blossom.
When a match is added to an alternating tree, the complexity is $O(m)$ from rescheduling the nodes, which exceeds the $O(q)$ cost of updating the alternating tree structure.
Finally, there is the cost of growing and shrinking regions.
In each stage, a node can only be involved in $O(1)$ ARRIVE or LEAVE flooder events: once a region is growing, it (or its topmost blossom ancestor) continues to grow until the next augmentation.
The cost of rescheduling nodes due to ARRIVE or LEAVE events in each stage is therefore $O(m)$.
We also remind the reader that the tracker has constant-time complexity per operation (since the radix heap has constant-time insert and extract-min operations), so it does not contribute to the overall complexity.
Therefore, in each stage the worst-case running time is dominated by the $O(nq^2 + mq)$ cost associated with up to $q$ blossoms forming or shattering.
There are $q$ stages, leading to a $O(nq^3 + mq^2)$ worst-case complexity.

\section{Handling negative edge weights}\label{app:negative_edge_weights}

Recall that an edge weight $\mathbf{w}[i]=\log((1-\mathbf{p}[i])/\mathbf{p}[i])$ can become negative since we can have $\mathbf{p}[i]>0.5$.
It is therefore necessary to to handle negative edge weights to decode correctly for these error models.
For example, consider the distance three repetition code check matrix
\begin{equation}
H=
\begin{pmatrix}
1 & 1 & 0\\
0 & 1 & 1\\
1 & 0 & 1
\end{pmatrix}
\end{equation}
with prior distribution $\mathbf{p}=(0.9, 0.9, 0.9)$ and an error $\mathbf{e}=(0, 1, 1)$ with syndrome $\mathbf{s}=H\mathbf{e}=(1, 0, 1)$.
The two errors consistent with the syndrome are $(0, 1, 1)$, which has prior probability $0.1\times 0.9^2=0.081$, and $(1, 0, 0)$, which has prior probability $0.9\times 0.1^2=0.009$.
Recall that MWPM decoding uses a weights vector $\mathbf{w}\in\mathbb{R}^3$ and finds a correction $\mathbf{c}\in\mathbb{F}_2^3$ satisfying $H\mathbf{c}=\mathbf{s}$ of minimal weight $\sum_i \mathbf{w}[i]\mathbf{c}[i]$.
Therefore, it is important that the edge weights $\mathbf{w}[i]=\log((1-\mathbf{p}[i])/\mathbf{p}[i])$ are indeed negative here, as this leads the decoder to predict the more probable (albeit higher \textit{hamming} weight) $\mathbf{c}=(0, 1, 1)$ instead of the incorrect $\mathbf{c}=(1, 0, 0)$.

We now show how negative edge weights can be handled for the MWPM decoding problem for some check matrix $H\in \mathbb{F}_2^{n\times m}$ with weights vector $\mathbf{w}\in\mathbb{R}^m$ and an error $\mathbf{e}\in\mathbb{F}_2^m$ with syndrome $\mathbf{s}=H\mathbf{e}\in\mathbb{F}_2^n$.
Even though sparse blossom only handles non-negative edge weights, we can still perform MWPM decoding when there are negative edge weights using the following procedure, which uses sparse blossom as a subroutine:
\begin{enumerate}
    \item Define $\mathbf{b}\in\mathbb{F}_2^m$ such that $\mathbf{b}[i]=1$ if $\mathbf{w}[i]<0$ and $\mathbf{b}[i]=0$ otherwise.
    \item From $\mathbf{b}$, define adjusted edge weights $\mathbf{v}\in\mathbb{R}^m$ where $\mathbf{v}[i]\coloneqq (1-2\mathbf{b}[i])\mathbf{w}[i]$, as well as the adjusted syndrome $\mathbf{s}^\prime = \mathbf{s}+H\mathbf{b}$.
    \item Use sparse blossom to find a correction $\mathbf{c}^\prime$ satisfying $H\mathbf{c}^\prime=\mathbf{s}^\prime$ of minimal adjusted weight $\sum_i\mathbf{v}[i]\mathbf{c}^\prime[i]$. Note that the definition of $\mathbf{v}$ guarantees that every element $\mathbf{v}[i]$ is non-negative.
    \item Return the correction $\mathbf{c}\coloneqq\mathbf{c}^\prime + \mathbf{b}$, which is guaranteed to satisfy $H\mathbf{c}=\mathbf{s}$ with minimal total weight $\sum_i\mathbf{w}[i]\mathbf{c}[i]$.
\end{enumerate}

We can verify the correctness of this procedure as follows.
Firstly, note that $\mathbf{c}^\prime$ satisfies $H\mathbf{c}^\prime=\mathbf{s}^\prime$ if and only if $\mathbf{c}$ satisfies $H\mathbf{c}=\mathbf{s}$.
Secondly, note that 
\begin{equation}
    \sum_i \mathbf{w}[i]\mathbf{c}[i]=\sum_i(\mathbf{w}[i]\mathbf{c}^\prime [i] + \mathbf{w}[i]\mathbf{b}[i] - 2\mathbf{w}[i]\mathbf{b}[i]\mathbf{c}^\prime[i])=\sum_i \mathbf{v}[i]\mathbf{c}^\prime [i] + \sum_i \mathbf{w}[i]\mathbf{b}[i].
\end{equation}
Here, the first equality uses $\mathbf{c}\coloneqq\mathbf{c}^\prime + \mathbf{b}$ and the term $- 2\mathbf{w}[i]\mathbf{b}[i]\mathbf{c}^\prime[i]$ accounts for the fact that the sum $\mathbf{c}^\prime + \mathbf{b}$ is taken over the binary field $\mathbb{F}_2$, so any bit set in both $\mathbf{c}^\prime$ and $\mathbf{b}$ should not have its corresponding weight $\mathbf{w}[i]$ contribute to the sum.
Therefore, if we find a $\mathbf{c}^\prime$ of minimal adjusted weight $\sum_i \mathbf{v}[i]\mathbf{c}^\prime [i]$ satisfying $H\mathbf{c}^\prime=\mathbf{s}^\prime$, it is guaranteed that the correction $\mathbf{c}\coloneqq\mathbf{c}^\prime + \mathbf{b}$ has minimal weight $\sum_i \mathbf{w}[i]\mathbf{c}[i]$ satisfying $H\mathbf{c}=\mathbf{s}$.
Intuitively, wherever we have an error mechanism with high error probability ($>50\%$), we are re-framing the decoding problem to instead predict if the error mechanism \textit{didn't} occur.
The handling of negative edge weights was also discussed in \cite{higgott2022pymatching}, as well as in Corollary 12.12 of \cite{korte2011combinatorial}, where what \citeauthor{korte2011combinatorial} refer to as a minimum-weight $T$-join is equivalent to what we call a MWEM.

\section{Example of detectors and observables for a repetition code}
\label{app:repetition_code_example}

\begin{figure}
    \centering
    \includegraphics[width=0.9\columnwidth]{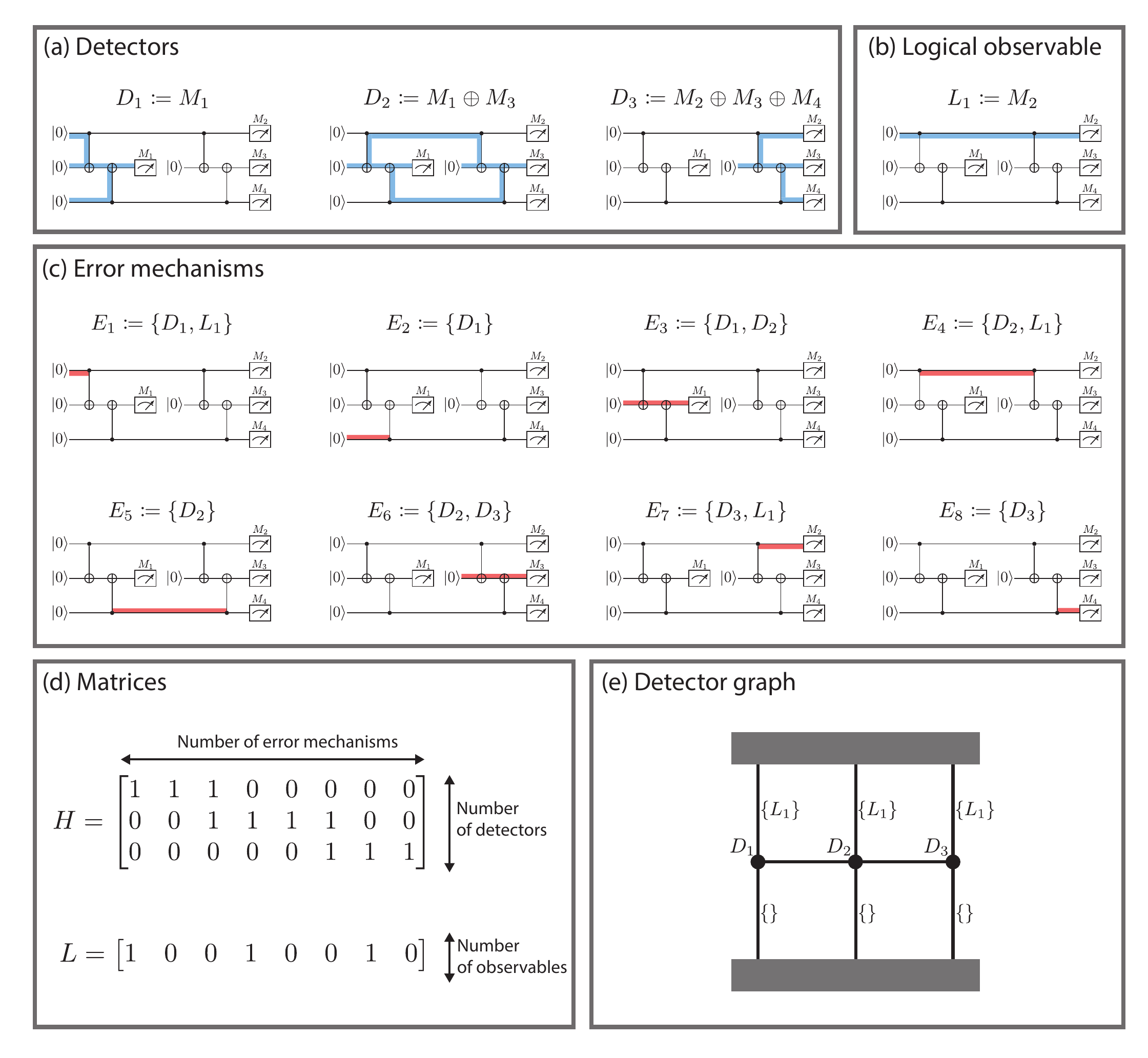}
    \caption{Representations of detectors, logical observables and error mechanisms for the circuit corresponding to a [[2,1,2]] repetition code memory experiment with two rounds of syndrome extraction. We use a bit-flip code (with stabilizer group $\langle ZZ\rangle$) and implement transversal initialisation and measurement in the $Z$ basis. (a) The three detectors in the circuit. The blue highlighted regions are the corresponding $Z$-type detecting regions~\cite{mcewen2023relaxing} (an error within this region which anti-commutes with its type will cause the corresponding detector to flip). (b) A logical $Z$ observable. Here, the blue highlighted region is the $Z$-type sensitivity region corresponding to the logical $Z$ observable - errors that anti-commute with $Z$ in this region will flip the outcome of the corresponding logical measurement $L_1$. (c) Given a stochastic Pauli noise model in the circuit, we can characterise errors based on the set of detectors and logical observables they flip. For a standard circuit-level depolarising noise model, there are eight different classes of error mechanism in this circuit, when classified this way. For each error mechanism, we highlight in red a region of the circuit where a single-qubit $X$ error would flip the same detectors and observables. Note that these single-qubit $X$ errors are just \textit{examples} of Pauli errors contributing to the error mechanisms; for example, another Pauli error contributing to $E_1$ would be a two-qubit $YX$ error after the first CNOT. (d) It is sometimes convenient to describe the detectors, observables and error mechanisms in terms of a detector check matrix $H$ and observable matrix $L$. Each column in $H$ or $L$ corresponds to an error mechanism. Each row in $H$ corresponds to a detector, with non-zero elements denoting the error mechanisms that flip the detector. Similarly, each row in $L$ corresponds to a logical observable, with non-zero elements denoting the error mechanisms that flip the observable. (e) If $H$ has column weight at most two, we can represent it with a detector graph $\mathcal{G}=(\mathcal{V},\mathcal{E})$. Each node $u\in\mathcal{V}$ corresponds to a detector. Each edge $(u,v)\in \mathcal{E}$ corresponds to an error mechanism that flips $u$ and $v$, each half-edge $(u,)\in\mathcal{E}$ corresponds to an error mechanism that flips just $u$, denoted here by an edge connected to the boundary. Each edge $e\in\mathcal{E}$ is assigned the set of logical observables $l(e)$ it flips, as well as a weight $w(e)\coloneqq \log((1-p_e)/p_e)$ (the weight is not shown in the diagram) where $p_e$ is the probability that the corresponding error mechanism occurs in the noise model. Note that here there are two parallel half-edges adjacent to each node; this is a symptom of the fact that the code has distance 2, and therefore has distinct error mechanisms that flip the same set of detectors but different sets of logical observables. In this instance, we can just keep the edge with the lower weight (since our implementation of sparse blossom does not directly handle parallel edges).}
    \label{fig:repetition_code_detectors}
\end{figure}

In \fig{repetition_code_detectors} we show the detectors, observables, matrices $H$ and $L$ and the detector graph $\mathcal{G}$ for the circuit corresponding to a memory experiment using a [[2,1,2]] bit-flip repetition code with two rounds of syndrome extraction.

Note that for a surface code memory experiment, the circuit and detector graph are instead three-dimensional.
In this case, the sensitivity region corresponding to a logical $Z$ observable measurement forms a 2D sheet in spacetime.
We denote this logical observable measurement $L_Z$.
This observable $L_Z$ is included in the set $l(e)$ of an edge $e\in\mathcal{E}$ if the error mechanism associated with $e$ flips $L_Z$.
For this to happen, the edge $e$ must pierce the 2D sheet (sensitivity region) and have $Z$-type detectors (detectors that are parities of $Z$ measurements) at its endpoints.

\end{document}